\documentclass{article}

\usepackage{graphicx}
\usepackage{enumerate}
\usepackage{amsmath,amssymb}               
\usepackage[a4paper]{geometry}
\usepackage{graphics}
\usepackage{soul,color}
\usepackage{natbib}
\usepackage{rotating}
\usepackage{hyperref}
\usepackage{setspace}

\usepackage{xcolor}
\usepackage{subfigure}
\usepackage{multirow}
\usepackage[normalem]{ulem}
\usepackage{float}

\newcommand{\beq}{\begin{equation}}
	\newcommand{\eeq}{\end{equation}}
\newcommand{\bce}{\begin{center}}
	\newcommand{\ece}{\end{center}}

\def\real{\hbox{\rm\setbox1=\hbox{I}\copy1\kern-.45\wd1 R}}

\newcommand{\bbe}{\mbox{\boldmath{$\beta$}}}

\newcommand{\bbh}{\mbox{\boldmath{$\hat{\beta}$}}}

\newcommand{\bZ}{{\bf Z}}

\newcommand{\bz}{{\bf z}}

\newcommand{\sumi}{\sum_{i=1}^n}

\newcommand{\sd}{\, \begin{picture}(-1,1)(-1,-2)\circle*{2}\end{picture}\ }
\newcommand{\pss}[1]{{(#1)}}
\newcommand{\cP}{\mathcal{P}}

\newcommand{\ph}{\phantom{$-$}}

\begin{document}

%%%%%%%%%%%%%%%%%%%%%%%%%%%%%%%%%%%%%%%%%%%%%%%%%%%%%%%%%%%%%%%%%
%%% Preamble
\vspace*{1cm}
\begin{center}
\Large
{\bf Revisiting the Cumulative Incidence Function \\ With Competing Risks Data} \\
\vspace*{8mm}
\large
David M.\ Zucker \\
Department of Statistics and Data Science, \\
The Hebrew University of Jerusalem, Jerusalem, Israel \\
david.zucker@mail.huji.ac.il \\
\vspace*{5mm}
and \\
\vspace*{5mm}
Malka Gorfine \\
Department of Statistics and Operations Research, \\
Tel Aviv University, Tel Aviv, Israel \\
gorfinem@tauex.tau.ac.il
\end{center}
\vspace*{15mm}
	
%%%%%%%%%%%%%%%%%%%%%%%%%%%%%%%%%%%%%%%%%%%%%%%%%%%%%%%%%%%%%%%%%
%%% Abstract
\begin{center}
Abstract
\end{center}
We consider estimation of the cumulative incidence function (CIF) in the competing risks Cox model. We study
three methods. Methods 1 and 2 are existing methods while Method 3 is a newly-proposed method. Method 3
is constructed so that the sum of the CIF's across all event types at the last observed event time is
guaranteed, assuming no ties, to be equal to 1. The performance of the methods is examined in a simulation
study, and the methods are illustrated on a data example from the field of computer code comprehension.
The newly-proposed Method 3 exhibits performance comparable to
that of Methods 1 and 2 in terms of bias, variance, and confidence interval coverage rates. Thus, with our
newly-proposed estimator, the advantage of having the end-of-study total CIF equal to 1 is achieved
with no price to be paid in terms of performance.

Keywords: Competing events; Computer program comprehension; Cox regression; Prediction; Survival analysis

%%%%%%%%%%%%%%%%%%%%%%%%%%%%%%%%%%%%%%%%%%%%%%%%%%%%%%%%%%%%%%%%%	
%%%%%%%%%%%%%%%%%%%%%%%%%%%%%%%%%%%%%%%%%%%%%%%%%%%%%%%%%%%%%%%%%	

\newpage
		
\section{Introduction}\label{sec1}

Competing risks arise when individuals are susceptible to several types of event and can experience at most one event.
Analysis of time to event without distinguishing between the different event types often yields an inadequate
picture of the data \citep[page 249]{kalbfleisch2011statistical}. There is a vast literature on competing risks, and the topic remains an active area of research.

This paper is motivated by the recently conducted experiment of \cite{ajami2019syntax} in the area of computer
program comprehension. Their goal was to measure how different syntactic and other factors influence code complexity and comprehension. To reach many subjects and obtain accurate measurements, they implemented
a website for the experiment designed based on some gamification principles, for details see \cite{ajami2019syntax}. 
The design consists of 40 code snippets, with each participant asked to interpret a subset of 11--14 snippets, presented in random order. The outcomes were time to answer and the accuracy of the snippet interpretation, i.e. correct or incorrect.
Thus, correct response and incorrect response were competing events. Out of the 2761 recorded trials, only 27 (0.98\%) of them ended in right censoring (i.e. no answer was provided after a certain period of time and the participant gave up). Because of this extremely low censoring rate, the censored individuals were excluded from the analysis and the data to be analyzed were free of censoring. 

Modeling based on \textit{cause-specific hazard} functions is a popular and useful approach for handling competing events  \citep[Section 3.2]{putter2007tutorial}. If we let $T$ denote the time to event and $D$ denote the type of event,
the cause-specific hazard $\lambda_j(t|{\bf z})$ for event type $j$, $j=1, \ldots, J$, for an individual with
covariate vector $\bz$ is defined as
\begin{equation*}
\lambda_j(t|{\bf z}) = \lim_{\epsilon \downarrow 0} \epsilon^{-1} P(T \in [t, t+\epsilon), D=j|\bz, T \geq t) \, .
\end{equation*}
This quantity represents the instantaneous incidence rate of cause $j$,  
given that the individual was free of any event up to time $t$. 
%Clearly,  $\lambda_j(t|{\bf z}) = f_j(t|{\b z})/S(t-|{\bf z})$, where $f_j(t|{\b z})$ and $S(t|{\bf z})$ are the conditional cause-$j$ density and overall survival, respectively. DZ - THIS IS NOT REALLY CORRECT.
Useful functions for prediction are the cause-specific cumulative incidence functions (CIFs), defined as
\begin{equation*}
F_j(t|\bz) = P(T \leq t, D=j|\bz) = \int_0^t S(u-|\bz) \lambda_j(u|\bz)  \, du \, ,
\end{equation*}
where
\begin{equation*}
S(t|\bz) = P(T > t|\bz) = \exp\left\{-\sum_{m=1}^J \Lambda_m(t|\bz)\right\}
\end{equation*}
with
\begin{equation*}
\Lambda_j(t|{\bz}) = \int_0^t \lambda_j(u|{\bz}) \, du \, . 
\end{equation*}
In the case of competing risks with no covariates, the CIF of each event type is usually estimated by the Aalen-Johansen estimator \citep{aalen1978empirical}. It can be shown that when the last observed time is an event time and there are no ties, 
the sum of the Aalen-Johansen estimators of the CIFs over all event types evaluated at the last event time is equal to 1.
In the presence of covariates, however, the situation is more complicated. 

In this paper, we study three methods for estimating the CIFs with covariates under a Cox-type model for the cause-specific
hazards. We refer to these methods as Method 1, Method 2, and Method 3. Methods 1 and 2 are existing estimators, while Method 3
is a newly-proposed estimator. The new method is
constructed so as to guarantee, in the absence of ties, that the sum of the CIF's across all event 
types at the last observed event time is equal to 1. Section 2 presents the methods, Section 3 presents
a simulation study comparing the methods, Section 4 presents an application of the methods
to the Ajami et al.\ program comprehension study, and Section 5 presents a short discussion.
R code for performing the simulations and data analysis in this paper is posted on the following webpage:

\url{https://github.com/david-zucker/cumulative-incidence-function.git}

The data used in the example are posted on the following webpage:

\url{https://github.com/shulamyt/break-the-code/tree/icpc17}

%%%%%%%%%%%%%%%%%%%%%%%%%%%%%%%%%%%%%%%%%%%%%%%%%%%%%%%%%%%%%%%%%	
%%%%%%%%%%%%%%%%%%%%%%%%%%%%%%%%%%%%%%%%%%%%%%%%%%%%%%%%%%%%%%%%%	

\section{Methods Considered}

To set the stage, we first consider the standard setup of ordinary survival data analyzed using the Cox model \citep{cox1972regression}.
For each individual $i$, we denote by $X_i$ the observed follow-up time on individual $i$ until the occurrence of
an event or censoring, and we set $D_i$ equal to 1 if individual $i$ experienced an event and equal to 0
if individual $i$ was censored. 
We define $N_i(t) = D_i I(X_i \leq t)$ and $Y_i(t) = I(X_i \geq t)$.
We assume that the event time has a continuous distribution, so that the probability of tied event times is 0.

A common estimator of the survival function $S(t|\bz)$ is given, as in Section 8.8 of \cite{klein2003survival}, by
\begin{equation}
\hat{S}^\pss{1}(t|\bz) = \exp \left\{-\hat{\theta}(\bz) \hat{\Lambda}_0(t) \right\}
\label{s_one}
\end{equation}
where 
$\hat{\theta}(\bz) = \exp\left\{ {\bbh}^T \bz \right\}$ and
\begin{equation*}
\hat{\Lambda}_0(t) = \sumi \int_0^\tau \left\{ \sum_{r=1}^n Y_r(t) \hat{\theta}(\bZ_r) \right\}^{-1} dN_i(t)
\end{equation*}
is the Breslow estimator of the  cumulative baseline hazard function.
An alternate estimator, given by Eqn.\ (7.2.34) of \cite{andersen2012statistical}, is
\begin{equation}
\hat{S}^\pss{2}(t|\bz) = \mathcal{P}_0^t \left\{ 1 - \hat{\theta}(\bz) d\hat{\Lambda}_0(s) \right\}
= \prod_{k=1}^{K(t)} \left\{ 1 - \hat{\theta}(\bz) \Delta \hat{\Lambda}_0\left(T_{\pss{k}}\right) \right\}
\label{s_two}
\end{equation}
where $\mathcal{P}_0^t$ denotes the product integral, $T_{\pss{k}}$ denotes the $k$-th ordered event time,
$K(t)$ denotes the number of event times in the interval $[0,t]$, and $\Delta \hat{\Lambda}_0(s) = \hat{\Lambda}_0(s)
- \hat{\Lambda}_0(s-)$. \cite{kalbfleisch2011statistical}, in Section 4.3, present  another possible estimator,
\begin{equation}
\hat{S}^\pss{3}(t|\bz) = \left( \prod_{k=1}^{K(t)} \hat{\alpha}_k \right)^{\hat{\theta}(\bz)}
\label{s_three}
\end{equation}
where, in the absence of tied survival times, $\hat{\alpha}_k$ is given by
\begin{equation*}
\hat{\alpha}_k = \left\{ 1 - \frac{\hat{\theta}\left(\bZ_{I(k)}\right)}
{\sum_{r=1}^n Y_r(T_{\pss{k}})\hat{\theta}(\bZ_r)} \right\}^{1/\hat{\theta}\left(\bZ_{I(k)}\right)} \, .
\end{equation*}
Here, $I(k)$ is the index of the individual who had the event at time $T_{\pss{k}}$.
Kalbfleisch and Prentice derived this estimator using a nonparametric maximum likelihood argument. 

All three of the above estimators are presented in Section VII.2.3 of \cite{andersen2012statistical}.
If the end-of-study risk set is large, these three estimators are nearly identical, whereas
if the end-of-study risk set is small they can differ substantially. 
With $\hat{S}^\pss{2}(t|\bz)$,
the quantity $\{1 - \hat{\theta}(\bz) \Delta \hat{\Lambda}_0(T_{\pss{k}}) \}$ can
go negative; a simple fix is to set $\hat{S}^\pss{2}(t|\bz)$ to 0 when this occurs.
The estimator $\hat{S}^\pss{3}(t|\bz)$, like the univariate Kaplan-Meier survival function estimator,
drops to 0 if the last observation time is as an event time. Our proposed analogue to $\hat{S}^\pss{3}(t|\bz)$
in the competing risk case, presented below, is constructed so as to achieve an analogous property: that the
estimated total cumulative distribution function of the time to event, taken over all event types, is equal to 1
if the last observation time is an event time.

If we define 
\begin{equation}
\hat{\gamma}_k(\bz) = 1 - \hat{\alpha}_k^{\hat{\theta}(\bz)} 
= 1 - \left\{ 1 - \frac{\hat{\theta}(\bZ_{I(k)})}
{\sum_{r=1}^n Y_r(T_{\pss{k}})\hat{\theta}(\bZ_r)} \right\}^{\hat{\theta}(\bz)/\hat{\theta}(\bZ_{I(k)})}
\label{gam}
\end{equation}
and
\begin{equation*}
\hat{\Gamma}(t|\bz) = \int_0^t \left[ 1 -  \left\{ 1 - \frac{\sum_{i=1}^n \hat{\theta}(\bZ_i) dN_i(s)} 
{\sum_{i=1}^n Y_i(s) \hat{\theta}(\bZ_i)} \right\}^{\hat{\theta}(\bz)/\sum_{i=1}^n \hat{\theta}(\bZ_i) dN_i(s)}
\right] \, \sum_{i=1}^n dN_i(s)
\end{equation*}
we can write
\begin{equation*}
\hat{S}^\pss{3}(t|\bz) = \prod_{k=1}^{K(t)} \left\{1-\hat{\gamma}_k(\bz)\right\} 
= \mathcal{P}_0^t \, \left\{1-\Delta \hat{\Gamma}(t|\bz) \right\} \, .
\end{equation*}
Comparing this with (\ref{s_two}), we can draw an association between $\Delta \hat{\Gamma}(t|\bz)$
and $\hat{\theta}(\bz) \Delta \hat{\Lambda}_0(T_{\pss{k}})$. If $\sum_{i=1}^n Y_i(s) \hat{\theta}(\bZ_i)$
is large, the two quantities are approximately equal, as may be seen using the
approximations $\log (1-u) \doteq -u$ and $1-e^{-v} \doteq v$.

We now move to the competing risk setting. We let $T, D, \lambda_j(t|\bz)$, and $F_j(t|\bz)$ be defined
as in the introduction.
We again denote the $k$-th ordered event time by $T_{\pss{k}}$, and we denote the corresponding event type
by $D_{\pss{k}}$. 
A common approach to modeling competing risks data is to use a Cox-model form for the cause-specific hazard:
\begin{equation*}
\lambda_j(t|\bz) = \lambda_{0j}(t)\exp({\bbe}_j^T \bz) \hspace{0.5cm} j=1,\ldots,J \, .
\end{equation*}
It is well-known that $\bbe_j$ can be consistently estimated using the Cox partial likelihood with
event types other than $j$ handled as censoring \citep[Section 8.2.3]{kalbfleisch2011statistical}.

Corresponding to the three survival function estimators presented above for ordinary survival data,
we can define three estimators of the CIF. 
Define $\hat{\theta}_j(\bz) = \exp\left(\bbh_j^T \bz\right)$ and $\hat{\theta}_{ij} = \hat{\theta}_j(\bZ_i)$.
The analogue of (\ref{s_one}) is then
\begin{equation*}
\hat{F}_j^\pss{1}(t|\bz) 
= \int_0^t \exp\left\{ - \sum_{m=1}^J \hat{\Lambda}_m(s-|\bz)\right\} d\hat{\Lambda}_j(s|\bz)
\end{equation*}
where
\begin{equation*}
\hat{\Lambda}_j(s|\bz) = \hat{\theta}_j(\bz) \int_0^t A_j(u)^{-1} \sumi dN_{ij}(u)
\end{equation*}
with
\begin{equation*}
A_j(u) = \sumi Y_i(u)\hat{\theta}_{ij} \, .
\end{equation*}
The analogue of (\ref{s_two}), as given by Section 8.5.1 of \cite{beyersmann2014classical}  is
\begin{equation*}
\hat{F}_j^\pss{2}(t|z) 
= \int_0^t \hat{P}(s-|\bz) d\hat{\Lambda}_j(s|\bz)
\end{equation*}
with
\begin{equation*}
\hat{P}(t|\bz) = \cP_0^t \left\{ 1 - \sum_{j=1}^J d \hat{\Lambda}_j(s|\bz) \right\}_+
= \prod_{k=1}^{K(t)} \left\{1 - \sum_{j=1}^J \Delta  \hat{\Lambda}_j(T_\pss{k}|\bz) \right\}_+
\end{equation*}
where $a_+ = \max(a,0)$.

Our proposed analogue of (\ref{s_three}) is 
\begin{equation*}
\hat{F}_j^\pss{3}(t|\bz) 
= \sum_{k=1}^{K(t)} \left[ \prod_{r=1}^{k-1} \left\{ 1-\hat{\gamma}_{r \sd}(\bz)\right\} \right] \hat{\gamma}_{kj}(\bz) 
\end{equation*}
where, analogously to (\ref{gam}), we define
\begin{equation*}
\hat{\gamma}_{kj}(\bz) = 1 - \left\{ 1 - \frac{\theta_j\left(\bZ_{I(k)}\right)I(D_{\pss{k}}=j)}{A_j(T_{\pss{k}})}
\right\}^{\hat{\theta}(\bz)/\theta_j(\bZ_{I(k)})}
\end{equation*}
and we set $\hat{\gamma}_{k \sd}(\bz) = \sum_{j=1}^J \gamma_{kj}(\bz)$.

As in the ordinary survival case, if the end-of-study risk set is large, the three estimators, $\hat{F}_j^{(m)}$, $m=1,2,3$, are nearly identical, whereas
if the end-of-study risk set is small they can differ substantially. Our proposed analogue of (\ref{s_three}) does not
have the nonparametric maximum likelihood interpretation that (\ref{s_three}) has, but it is still a plausible estimator.

We define $F_{\sd}(t|\bz) = \sum_{j=1}^J F_j(t|\bz)$, which is the probability that an individual with
covariate vector $\bz$ experiences an event of some type during the interval $[0,t]$.
Correspondingly, for $m = 1, 2$, and 3, we define $\hat{F}_{\sd}^\pss{m}(t|\bz) = \sum_{j=1}^J \hat{F}_j^\pss{m}(t|\bz)$. 

It is an algebraic fact, which can be proved by induction, that for any $c_1, \ldots, c_K$ we have
\begin{equation*}
1 - \sum_{k=1}^K \left\{ \prod_{r=1}^{k-1} (1-c_r) \right\} c_k  = \prod_{r=1}^K (1-c_r)	
\end{equation*}
Thus, 
\begin{equation*}
1 - \hat{F}_{\sd}^\pss{3}\left(T_{\pss{K}}|\bz\right) = \prod_{r=1}^K \left\{1-\hat{\gamma}_{r \sd}(\bz)\right\} \, .	
\end{equation*}
Now, if $T_{\pss{K}}$ is the last observed follow-up time (i.e., the last observed follow-up time was an event),
then $A_j(T_{\pss{K}}) = \theta_j(\bZ_{I(K)})$, and so we have $\hat{\gamma}_{Kj}(\bz) = I(D_{I(K)} = j)$
and $\hat{\gamma}_{K \sd}(\bz)=1$. Thus, in this case, such as with uncensored data, we obtain
$1-\hat{F}_{\sd}^\pss{3}(T_{\pss{K}})=0$ and $\hat{F}_{\sd}^\pss{3}(T_{\pss{K}})=1$.
The estimators $\hat{F}_j^\pss{1}(t)$ and $\hat{F}_j^\pss{2}(t)$ do not have this property.
In fact, for these estimators, $\hat{F}_{\sd}^\pss{m}(T_{\pss{K}})$ can exceed 1.

We also considered a 95\% simultaneous confidence band of the form $\hat{F}^\pss{j}(t|\bz) \pm c_{j,0.95}$ (i.e., a
fixed-width band) for $F^\pss{j}(t|\bz)$ over $t \in [0,T_{\pss{K}}]$. We computed the critical value $c_{j,0.95}$ using the weighted bootstrap 
\citep[Section 8.2]{kosorok2007inference}.
For each bootstrap replication $m$, a set of weights $w_{mi}^\circ$ is generated as random draws from the
$Exp(1)$ distribution, normalized weights are computed as $w_{mi}^\circ / (n^{-1} \sum_{r=1}^n w_{mr}^\circ)$,
and the bootstrap estimate is computed by replacing quantities of the form
$\sum_{i=1}^n term_i$ by $\sum_{i=1}^n w_{mi} \, term_i$. The critical value $c_{j,0.95}$ is then taken to be
the 95th percentile across the bootstrap replications of the maximum absolute difference
between the bootstrap estimate and the estimate for the original data.

\section{Simulation Study}

We conducted a simulation study to compare the estimates $\hat{F}_j^\pss{m}(t|\bz), m = 1, 2, 3$. We considered
a setup with two competing risks, one with a high final CIF (65\%) and one with a low final CIF (35\%).
There was a single covariate $Z$, with distribution $U(-0.5,0.5)$.
We used a baseline hazard of the form
\begin{equation*}
\lambda_0(t) = \frac{\sigma p (t+a)^{p-1}}{1+b(t+a)^p}
\end{equation*}
with a corresponding cumulative baseline hazard of the form
\begin{equation*}
\Lambda_0(t) = \sigma b^{-1} \log (1+b(t+a)^p)
\end{equation*}
As $b$ tends to 0, we get a Weibull-type model with $\Lambda_0(t) = \sigma (t+a)^p$
and $\lambda_0(t) = \sigma p (t+a)^{p-1}$. We considered three shapes for 
the baseline hazard function, increasing ($a=0, b=0, p=3$), decreasing $(a=0.4, b=0, p=0.5)$ and up-and-down 
($a=0, b=0.75, p=3$). The parameter $\sigma$ was set so as to achieve CIF values approaching the desired final values
at about time $t=5$. The survival distributions were truncated at time $t=10$.
We ran simulations for a sample size of 75 with no censoring and for a sample size of 150 with 50\% censoring.
We took the regression coefficient
to be either $\log 3$ or $\log 6$ for both event types, so that the relative risk associated with
a 1 unit increase in the covariate value was either 3 or 6. The estimates of $F_j(t|z)$ were computed at
$z=-0.4$, $z=0$, and $z=0.4$.
In all simulations, 1,000 simulation replications were run,
and for each simulation replication the bootstrap confidence band procedures were carried out using 1,000 bootstrap
replications. Table \ref{tbl:sim-unif} summarizes the configurations studied under a uniformly distributed covariate. 

We also conducted a supplemental set of simulations where
the final CIF was again set to be about 65\% for the high CIF event and 35\% for the low CIF event,
but now the covariate had distribution $N(0,4)$. We considered the case of uncensored data with sample size 75.
We used the increasing hazard. The values for the regression coefficient were again 
either $\log 3$ or $\log 6$ for both event types. The estimates of $F_j(t|z)$ were computed at
$z=-1.68$, $z=0$, and $z=1.68$ ($-1.68$ and 1.68 are, respectively, the 20th and 80th percentiles of
the $N(0,4)$ distribution). 

Figures 1--5 present the results and Tables S.1--S.8 in the
Supporting Information present the same results in tabular form. 
Figures 1 and 3 shows the maximum mean bias of the CIF estimates
up to the 90th percentile of the last observed event time, the standard error of the estimates at the 90th percentile 
of the last observed event time
(the standard error generally increased over time), the empirical coverage rates of the 95\% confidence bands, and the 
half-width of the 95\% confidence bands. Figure 1 depicts the results of uniformly distributed covariate, and Figure 3
of normally distributed covariate. Methods 2 and 3 generally yielded a maximum bias of less than 0.01 whereas the 
bias with Method 1 tended to be higher.
The three estimates were comparable in terms of maximum standard error.
Regarding the confidence bands, Methods 1 and 2 often yielded low coverage rates, in some cases as low as 0.8. Method
3 tended to yield wider confidence bands but with proper coverage rates.

Figures 2 and 4 (for uniformly and normally distributed covariate, respectively) show, for the configurations
without censoring, various quantiles (1\%, 10\%, 50\%, 90\%, 99\%) of the total CIF (i.e., the CIF summed over the two risks) at the last event time. Tables S.9 and S.10 in the Supporting Information present the same results in tabular form.
Since $\hat{F}^{(3)}_{\sd}(T_{(K)}|\bz)=1$ by construction, this estimator is not included in these figures and tables. Although without censoring, the estimators should be exactly 1, we see that  $\hat{F}^{(1)}_{\sd}$ and $\hat{F}^{(2)}_{\sd}$ could deviate from 1, sometimes substantially, and can be less than or greater than 1. The magnitude of the deviation from 1 was usually
larger with $\hat{F}^{(1)}_{\sd}$ than with $\hat{F}^{(2)}_{\sd}$.

\section{Real Data Example -- Program Comprehension}
Most of a software engineer's time is spent reading codes. Sometimes this is their own code, while most of the time it is someone else's code. This reading is often named \textit{program comprehension}. Being good at program comprehension is a critical skill but it is notoriously hard and time consuming.  Surprisingly, there has been relatively little empirical
work on how program structures effect comprehension. Recently, Ajami et al. \cite{ajami2019syntax} used an experimental platform fashioned as an online game-like environment to measure how quickly and accurately 222 professional
programmers can interpret code snippets with similar functionality but different structures. Their goal was to measure how different syntactic and other factors influence code complexity and comprehension. For example, what is the effect of control structures on code complexity?
Is the complexity of an \texttt{if} the same as that of a \texttt{for}? For the complete list of their research questions, see \cite{ajami2019syntax}. 

The following summary of the design description is based on \cite{ajami2019syntax}.
The experiment was conducted by showing participants short code snippets which they needed to interpret. All code segments checked whether a number is in a set of non-overlapping ranges. The design consists of 40 code snippets: 12 each with 3-range and 4-range versions, 9 with 2-range versions, and 7 special loop cases. Table \ref{tbl:snippets} provides a concise description of the snippets. In a pilot study they found that reading 40 snippets is too much for a single participant to perform, so a subset of snippets was selected to each participant. The selection was done efficiently in terms of including pairs or sets of snippets that are meaningful to compare to each other. The total number of snippets presented to each subject was between 11 and 14, presented in random order. The outcomes were the time to answer and accuracy of the response (correct/incorrect). To reach many subjects and achieve accurate measurements, they implemented a website for the
experiment, designed based on some gamification principles, for details see their paper. At the beginning of the experiment,  
a popup was opened with a demographic questionnaire and details on education and experience. The choice of a test
plan did not depend on experience or any other demographic information of the participant. Afterward, an example screen was displayed, showing how the experimental screen looks and explaining the ``game'' rules. When the actual experiment started, a code snippet was presented and the subject were supposed to type in the snippet's output.  

In the above setup, correct and incorrect response are competing events. Time was measured from displaying the code until the participant pressed the button to indicate he/she was done. Another outcome variable was the button the subject chose to click: either ``I think I made it'' or ``skip'', where skip indicates right censoring. However, skip was used only 27 times in total, out of 2761 recorded trials, so its effect is negligible and these 27 cases were excluded. Thus, the dataset does not involve censoring. If a participant decided to quit the experiment before completing the snippets set, the analysis consisted of the completed snippets. Another important issue is the order in which the snippets were presented to each participant, as the common framework behind the snippets may lead to learning effects. Therefore, one of the covariates in the following analysis is the snippet's place in the sequence of snippets (snippet order).

In total there were 1893 correct answers and 868 incorrect answers.  On average, a question was answered by 58.15 participants. 
To demonstrate the differences between the three CIF estimators and the advantage of $\hat{F}_j^{(3)}$, $j=1,2$, we show here the result of the snippet \texttt{lp3} (a snippet of ``special loop" type). The analysis is based on 69 players; 49 provided a correct answer and 20 provided an incorrect answer. The covariates included in the Cox regression analysis were the participant's age, sex and years of experience (YoE) and the snippet's order. Figures \ref{fig:plot103A} and \ref{fig:plot103B} display $\hat{F}_j^{(m)}(\cdot|\bz)$ and $\hat{F}^{(m)}_{\sd}(\cdot|\bz)$, $j=1,2$, $m=1,2,3$, for various $\bz$. The confidence bands were omitted for simplicity of presentation. Table \ref{tbl:snippets2} presents $\hat{F}_{\sd}^{(m)}(T_{(K)}|\bz)$, for $m=1,2$; $\hat{F}_{\sd}^{(3)}(T_{(K)}|\bz)=1$ in all cases and is thus omitted from the table. Evidently, there is a substantial learning effect during the experiment, since the chance of a correct answer increases with the snippet's order. Moreover,  $\hat{F}_{\sd}^{(m)}(T_{(K)}|\bz)$ for $m=1,2$, is sometimes above 1 and sometimes much less than 1, e.g. 0.87. Figure \ref{fig:plot103C} shows the CIFs of $\bz = (\mbox{snippet order}=1,\mbox{age}=35,\mbox{female},\mbox{YoE}=5)$ of the three methods with 95\% confidence bands. 

\section{Discussion}

In this paper, we have studied three methods for estimating the cumulative incidence function (CIF) in the
competing risks Cox model: two existing methods and a newly-proposed method. The new method is
constructed so as to guarantee, in the absence of ties, that the sum of the CIF's across all event 
types at the last observed event time is equal to 1. By constrast, with the existing methods,
the sum of the CIF's over all event types evaluated
at the last event time can be less than or greater than 1. In an extensive simulation study, we showed that the deviation 
from 1 can be substantial under small sample size. In a recent paper, \cite{austin2021} discussed the issue of
the total CIF exceeding 1 in the context of the Fine-Gray model, another competing risks model, and
pointed out that this is a problematic phenomenon. For the Cox cause-specific hazard model,
the phenomenon tends not to appear when there is a high percentage of censoring, but here we
have seen that for uncensored data the phenomenon can appear with Methods 1 and 2.
Method 3 avoids this undesirable phenomenon.
Our simulations showed further than
the newly-proposed Method 3 exhibits performance comparable to that of the existing methods
in terms of bias, variance, and confidence interval coverage rates. Thus, with our
newly-proposed estimator, the advantage of having the end-of-study total CIF equal to 1 is achieved
without paying a price in terms of performance.

\bibliographystyle{chicago}
\bibliography{literature}

\newpage

\begin{table}[H]
	\caption{\label{tbl:sim-unif} Summary of simulation configurations under $Z \sim U(-0.5,0.5)$.}
	\centering
	\begin{tabular}{cccccc}
		\hline	
	Scenario &  {Hazard Type} &    Sample size   & $\exp(\beta_j)$ &      $z$ &   Censoring rate \\
	\hline   
	1  &    increasing &   75   & 3  & -0.4  &    0.0  \\    
	2   &   increasing &  150   & 3  & -0.4  &   0.5  \\    
	3   &   increasing &   75   & 3  &  0.0  &   0.0  \\    
	4   &   increasing &  150   & 3  &  0.0   &   0.5 \\     
	5   &   increasing &   75   & 3  &  0.4   &   0.0  \\   
	6   &   increasing &  150   & 3  &  0.4   &   0.5  \\    
	7   &   increasing &   75   & 6  & -0.4   &   0.0  \\    
	8   &   increasing &  150   & 6  & -0.4   &   0.5  \\    
	9   &   increasing  &  75   & 6  &  0.0   &   0.0  \\    
	10  &    increasing &  150  &  6 &   0.0  &    0.5 \\     
	11   &   increasing &   75  &  6 &   0.4  &    0.0  \\   
	12   &   increasing &  150   & 6 &   0.4  &    0.5  \\    
	13   &   decreasing &   75   & 3  & -0.4  &    0.0  \\    
	14   &   decreasing  & 150   & 3  & -0.4  &    0.5  \\    
	15   &  decreasing   & 75    & 3  &  0.0  &    0.0  \\    
	16   &   decreasing  & 150   & 3  &  0.0  &    0.5  \\   
	17   &   decreasing  &  75   &3   & 0.4   &   0.0   \\   
	18   &   decreasing  & 150   & 3  &  0.4  &    0.5  \\   
	19   &   decreasing  &  75   & 6  & -0.4   &   0.0  \\    
	20   &   decreasing  & 150   & 6  & -0.4   &   0.5  \\    
	21   &   decreasing  &  75   & 6  &  0.0   &   0.0  \\    
	22   &  decreasing   &  150  &  6 &   0.0  &    0.5 \\  
	23   &   decreasing  &  75   & 6   & 0.4    &  0.0  \\    
	24   &  decreasing   &  150  &  6  &  0.4   &   0.5 \\    
	25   &  up-and-down  &  75    & 3   &-0.4    &  0.0  \\    
	26   &   up-and-down  & 150   & 3   &-0.4    &  0.5  \\    
	27   &   up-and-down  &  75   & 3   & 0.0    &  0.0 \\     
	28   &   up-and-down  & 150   & 3   & 0.0    &  0.5  \\    
	29   &   up-and-down  &  75   & 3   & 0.4    &  0.0  \\    
	30   &   up-and-down  & 150   & 3   & 0.4    &  0.5 \\     
	31   &   up-and-down  &  75   & 6   &-0.4    &  0.0  \\    
	32   &   up-and-down  & 150   & 6    &-0.4   &   0.5 \\     
	33   &   up-and-down  &  75   & 6    &0.0    &  0.0 \\     
	34   &   up-and-down  & 150   & 6    &0.0    &  0.5 \\     
	35   &   up-and-down  &  75   & 6    &0.4    &  0.0 \\     
	36   &   up-and-down  & 150   & 6    &0.4     & 0.5 \\     
		\hline
\end{tabular}
\end{table}		

\newpage

\begin{table}[H]
		\caption{\label{tbl:snippets} Summary of code snippets. All code segments check whether a number is in a set of non-overlapping ranges.
		If a code snippet is preceded by a number, it indicates the number of ranges.}
	\centering
	\begin{tabular}{|c|p{12cm}|}
		\hline
		Snippet  & Description \\
		\hline
		\hline
		as,bs,cs & Structure variants a,b,c - use if else with several conditions. \\
		\hline
		al,bl (b1l),cl & Logical variants a,b,c - use nested single condition if else. \\
		\hline
		an,an1,an2 & Variants of snippet al that use negation. \\
		\hline
		f* & For loop that uses simple arithmetic manipulations of the for loop iteration variable. \\
		\hline
		f[] & For loop that uses pointer array for the ranges' limits. \\
		\hline
		lp0$,\ldots,$lp6 & Special loops. The for loop statement is used differently from common practice. For example $i$ counting down instead of up. \\
		\hline
	\end{tabular}
\end{table}

\begin{table}[H]
		\caption{\label{tbl:snippets2} Analysis of code snippets \texttt{lp3}. YoE - years of experience. Since the data are free of censoring, the desired estimator of the marginal CIF at the last failure time is exactly 1,  but only Method 3 provides it.
			}
	\centering
	\begin{tabular}{cccc|cc}
		\hline	
	Snippet order & Age & Sex & YoE & $\hat{F}^{(1)}_{\sd}(T_{(K)}|\bz)$ &   $\hat{F}^{(2)}_{\sd}(T_{(K)}|\bz)$	 \\
		\hline	
  1 & 35 & female & 0 & 0.7969 & 0.7896 \\
  1 & 35 & male & 0 & 0.9321 & 0.9151 \\
 1 & 35 & female & 5 & 0.8750 & 0.8632 \\
 1 & 35 & male & 5 & 0.9834 & 0.9593 \\
 10 & 35 & female & 0 & 1.0423 & 1.0036 \\
 10 & 35 & male & 0 & 1.0385 & 1.0008 \\
 10 & 35 & female & 5 & 1.0415 & 1.0023 \\
 10 & 35 & male & 5 & 1.0350 & 1.0001 \\
		\hline
\end{tabular}
\end{table}

\begin{figure}[H]
	\centering
	\makebox{\includegraphics[width=15cm,height=16cm]{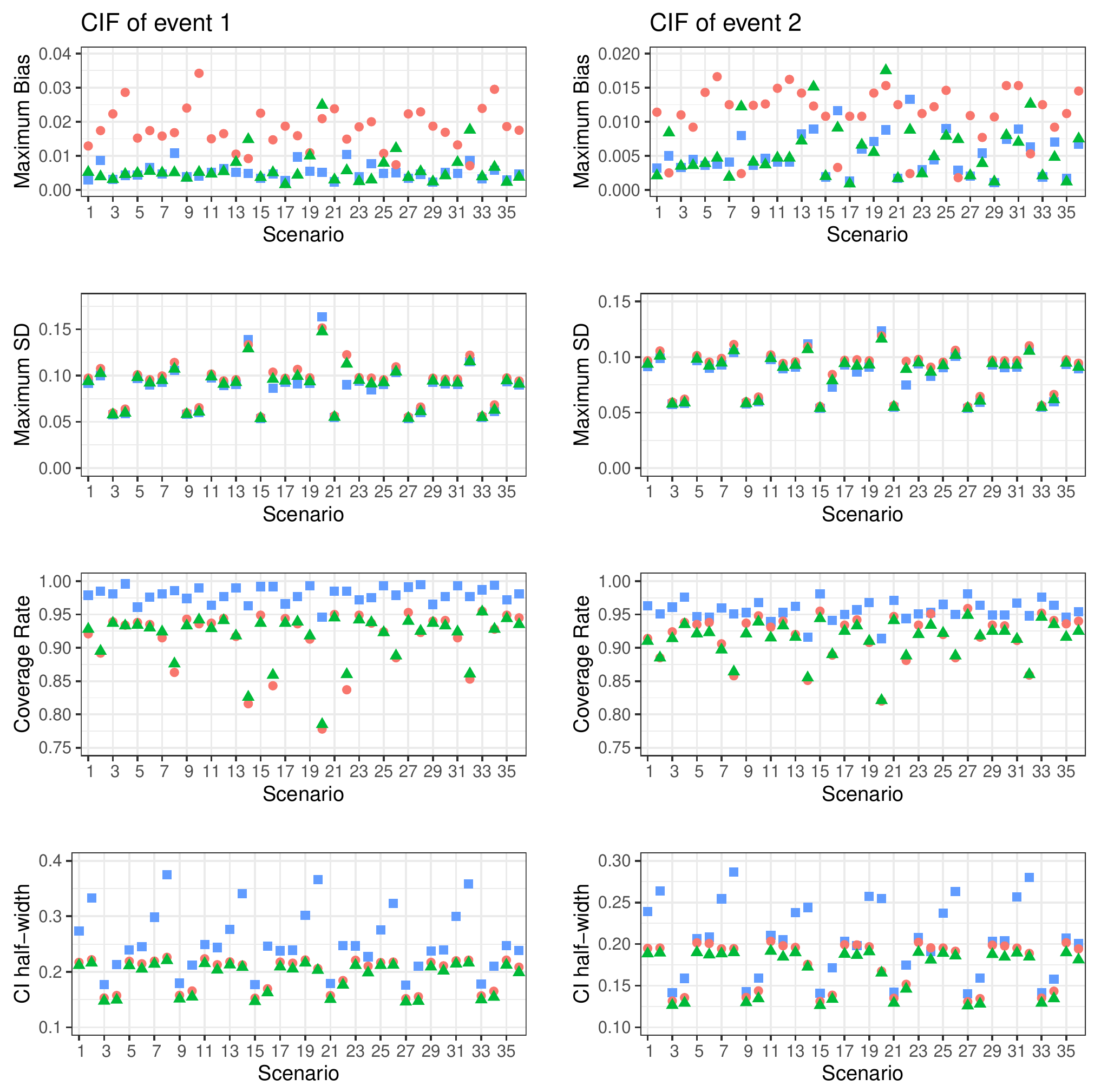}}
	\caption{\label{fig:unifbias} Simulation results with uniformly distributed covariate. $\hat{F}^{(1)}_j$ - red circle; $\hat{F}^{(2)}_j$ - green triangle; $\hat{F}^{(3)}_j$ - blue square. Line 1 - the maximum mean bias of the CIF estimates
		up to the 90th percentile of the last observed event time; line 2 - the standard error of the estimates at the 90th percentile of the last observed event time
		(the standard error generally increased over time); line 3 -  the empirical coverage rates of the 95\% confidence bands; and line 4 - half width of the 95\% confidence bands. See Table \ref{tbl:sim-unif} for the scenarios' configurations.}
\end{figure}

\begin{figure}[H] 
	\centering
	\makebox{\includegraphics[width=15cm,height=16cm]{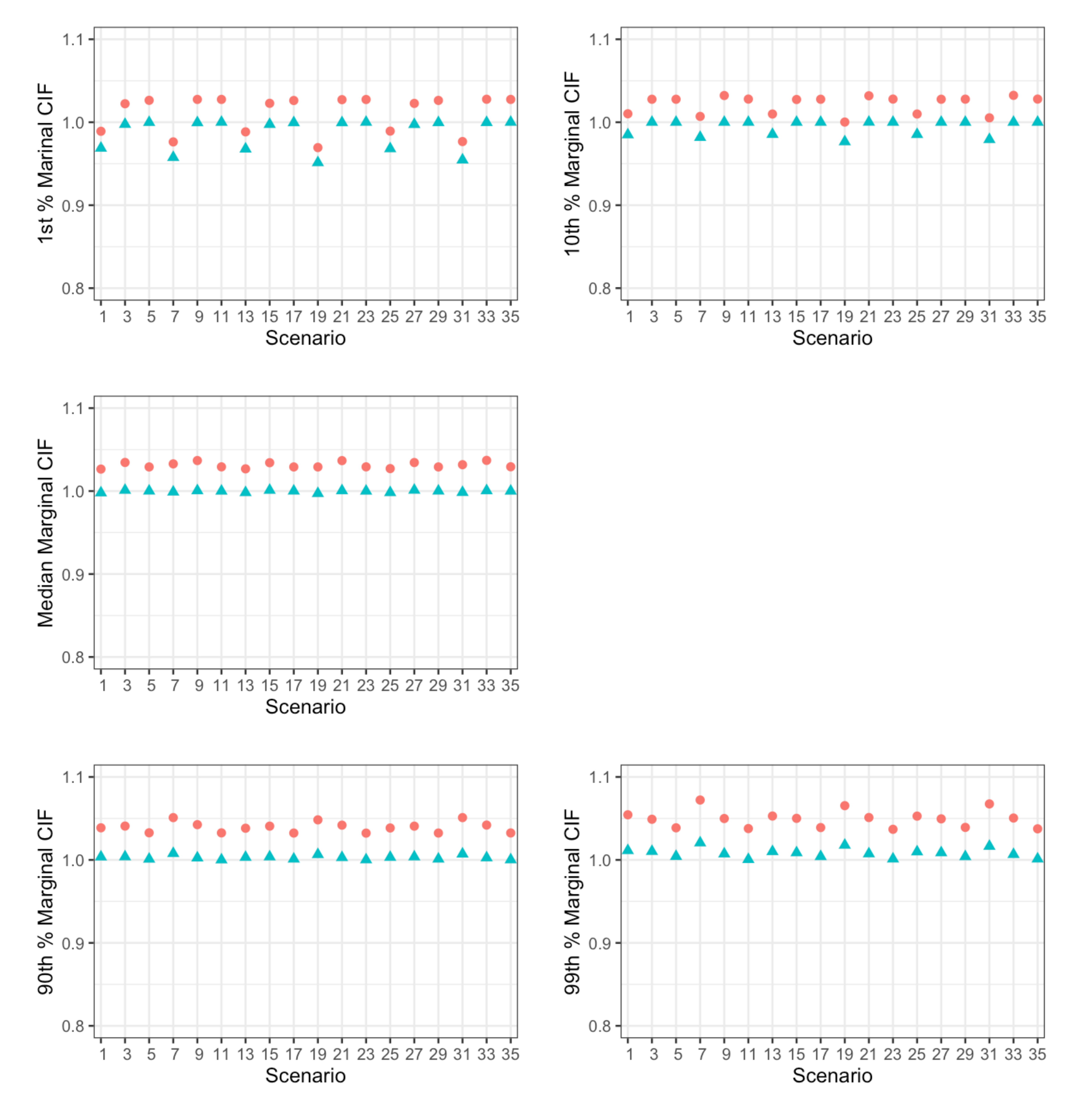}}
	\caption{\label{fig:unifperc} Simulation results with uniformly distributed covariate. $\hat{F}^{(1)}_j$ - red circle; $\hat{F}^{(2)}_j$ - green triangle.  Various quantiles (1\%, 10\%, 50\%, 90\%, 99\%) of the marginal CIF (i.e. the CIF summed over the two risks) at the last event time, for the configurations	without censoring. See Table \ref{tbl:sim-unif} for the scenarios' configurations.}
\end{figure}

\begin{figure}[H] 
	\centering
	\makebox{\includegraphics[width=15cm,height=16cm]{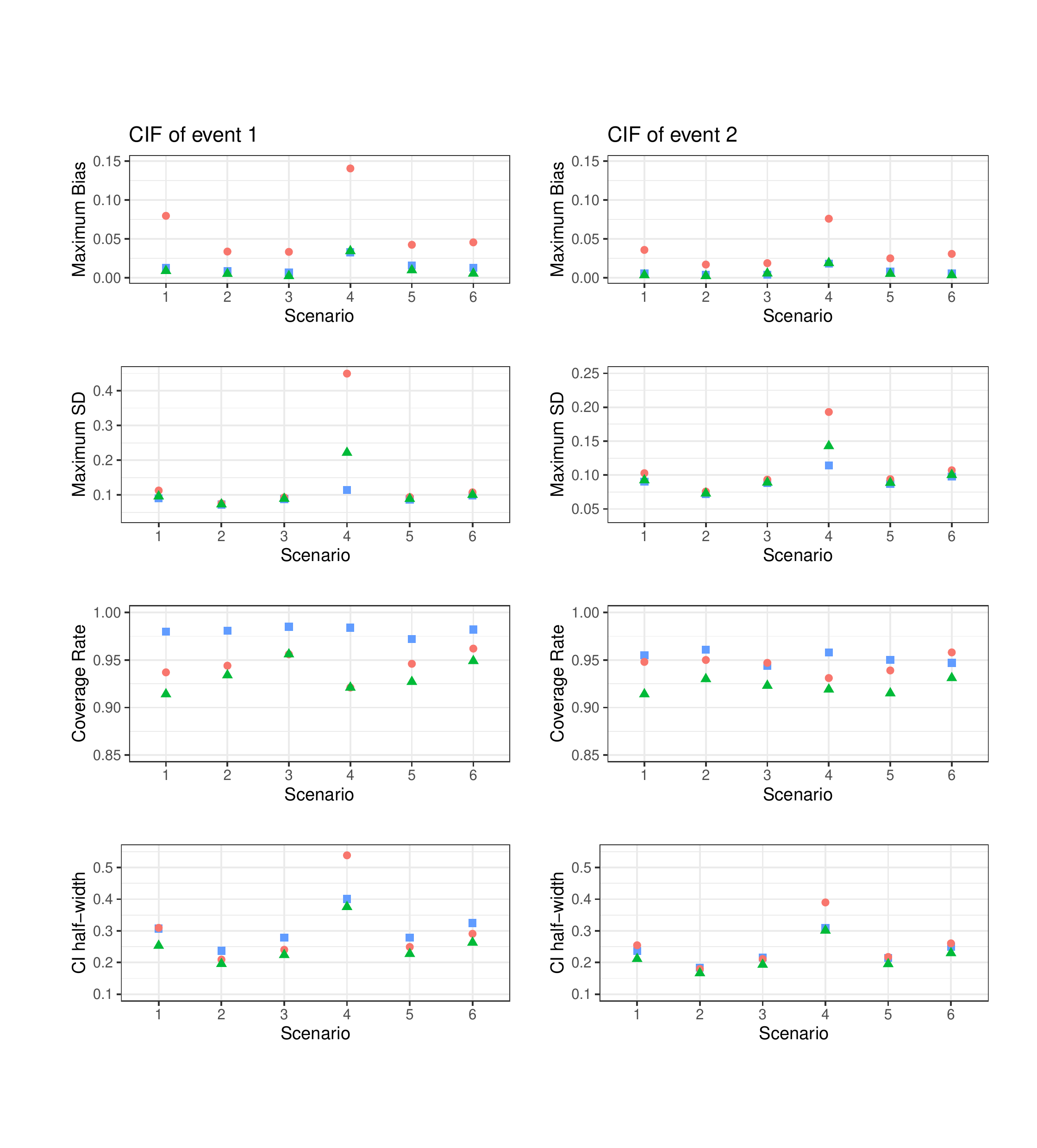}}
	\caption{\label{fig:normalbias} Simulation results with normally distributed covariate. $\hat{F}^{(1)}_j$ - red circle; $\hat{F}^{(2)}_j$ - green triangle; $\hat{F}^{(3)}_j$ - blue square. Line 1 - the maximum mean bias of the CIF estimates
		up to the 90th percentile of the last observed event time; line 2 - the standard error of the estimates at the 90th percentile of the last observed event time
		(the standard error generally increased over time); line 3 -  the empirical coverage rates of the 95\% confidence bands; and line 4 - half width of the 95\% confidence bands. $n=75$; no censoring; Scenarios 1 and 4 with $z=-1.68$; Scenarios 2 and 5 with $z=0$; Scenarios 3 and 6 with $z=1.68$. }
\end{figure}

\begin{figure}[H] 
	\centering
	\makebox{\includegraphics[width=15cm,height=16cm]{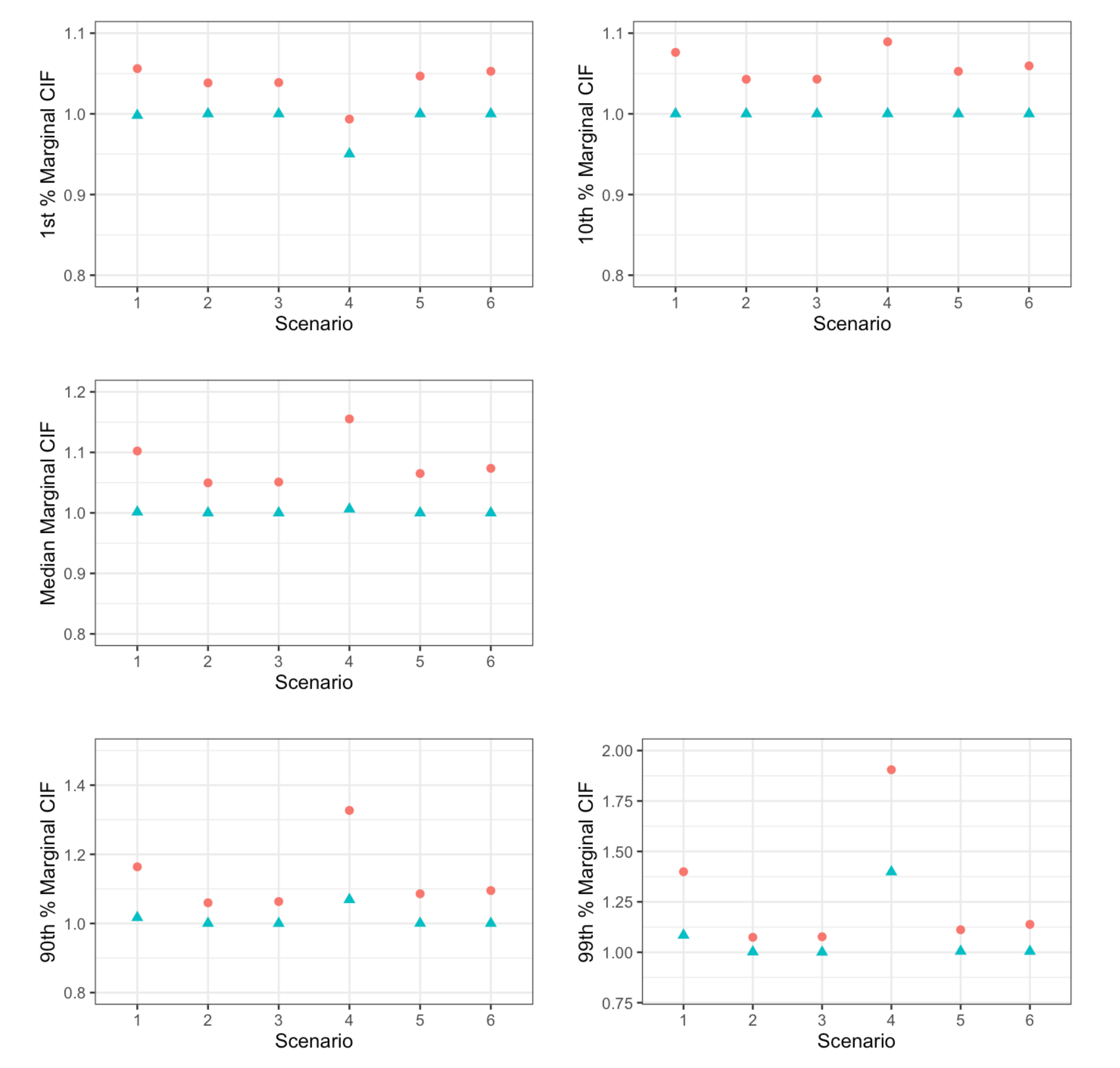}}
	\caption{\label{fig:normalperc} Simulation results with normally distributed covariate. $\hat{F}^{(1)}_j$ - red circle; $\hat{F}^{(2)}_j$ - green triangle.  Various quantiles (1\%, 10\%, 50\%, 90\%, 99\%) of the marginal CIF (i.e. the CIF summed over the two risks) at the last event time, for the configurations	without censoring. $n=75$; no censoring; Scenarios 1 and 4 with $z=-1.68$; Scenarios 2 and 5 with $z=0$; Scenarios 3 and 6 with $z=1.68$.}
\end{figure}

\begin{figure}[H]
	\centering
	\makebox{\includegraphics[width=15cm,height=16cm]{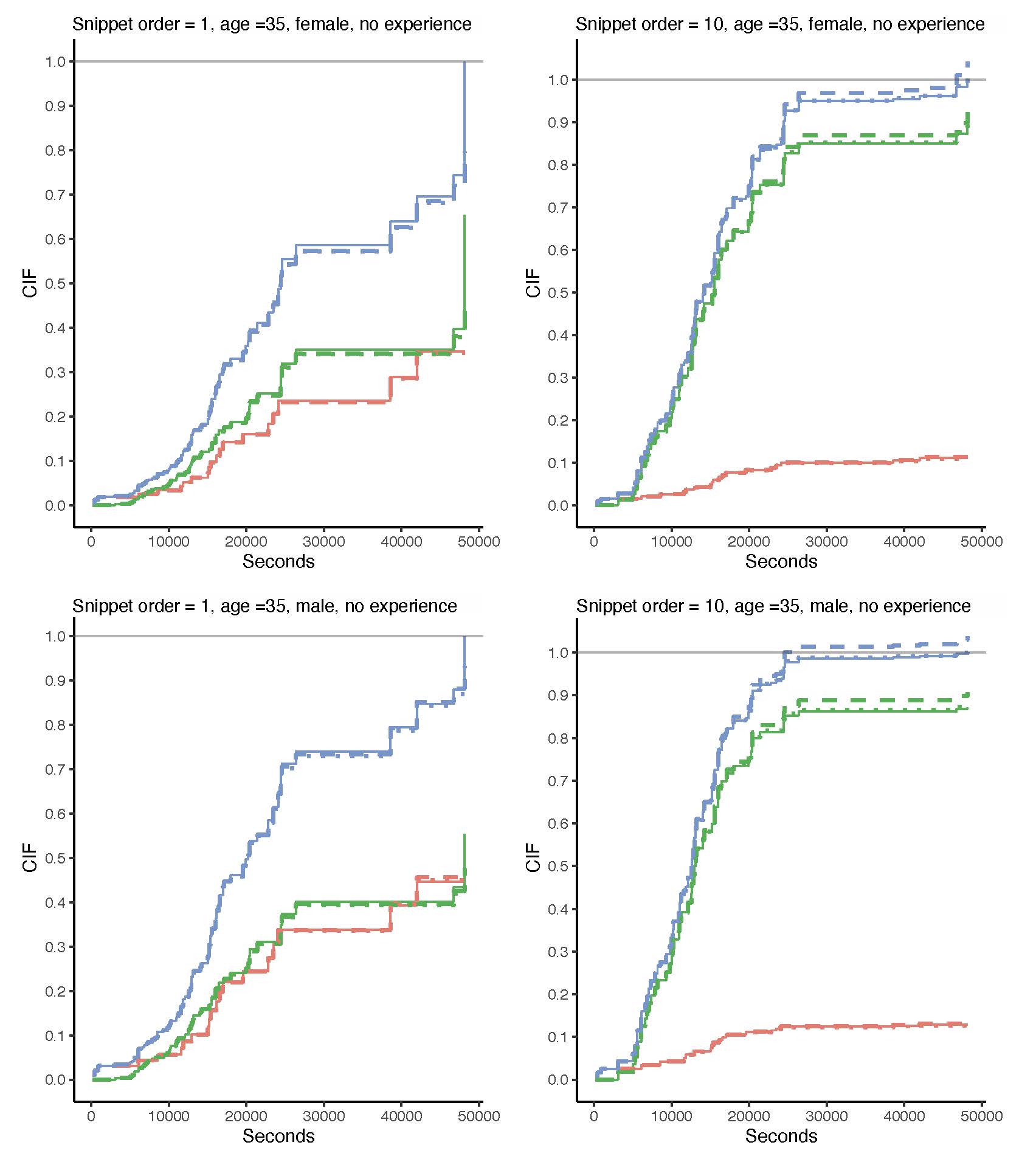}}
	\caption{\label{fig:plot103A} Data Analysis, Snippet \texttt{lp3}, less than a year of programming experience. $\hat{F}^{(1)}_j$ - dashed line; $\hat{F}^{(2)}_j$ - dotted line; $\hat{F}^{(3)}_j$ - continuous line. Red - incorrect answer; green - correct answer, blue - marginal probability.}
\end{figure}

\begin{figure}[H]
	\centering
	\makebox{\includegraphics[width=15cm,height=16cm]{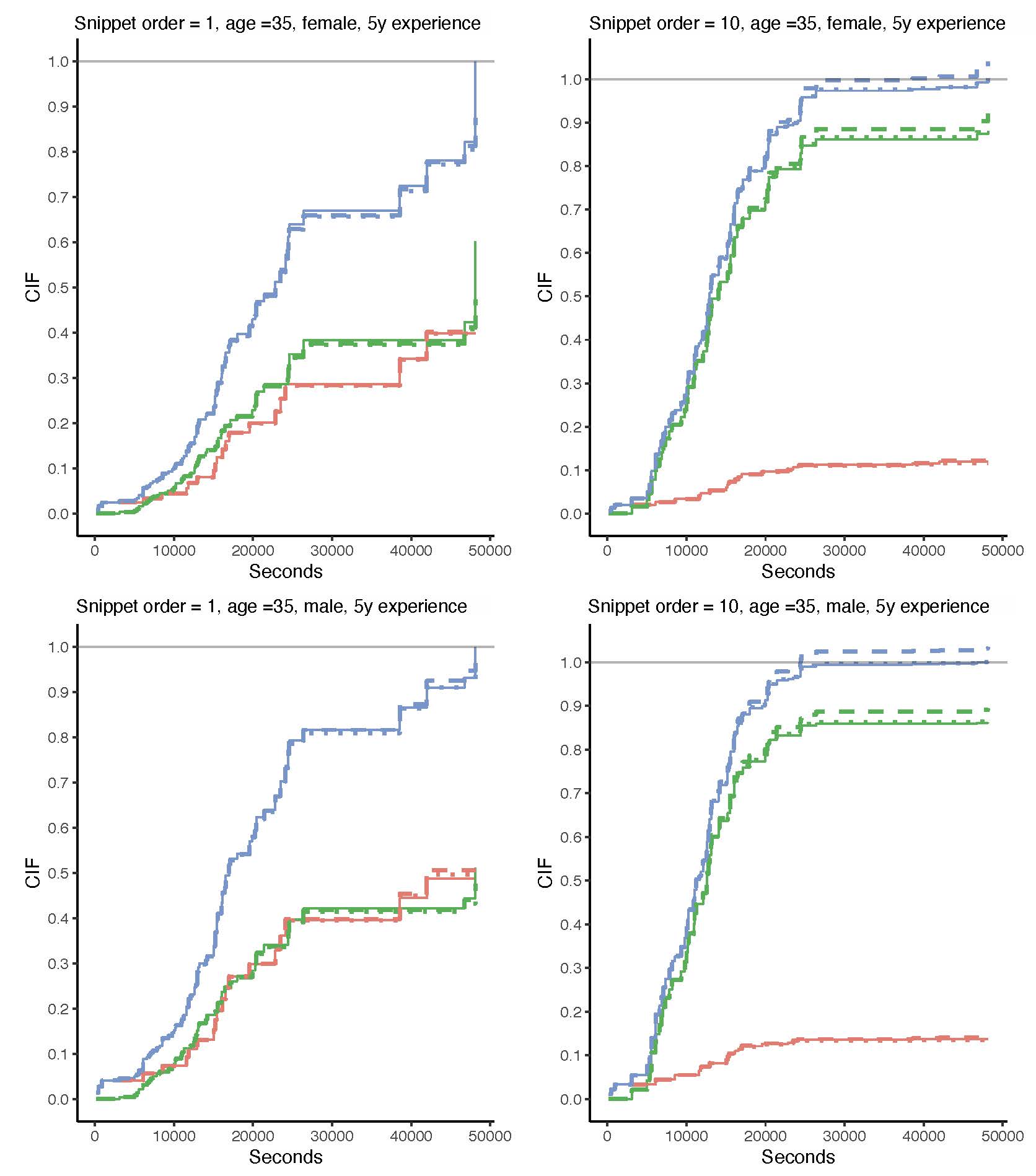}}
	\caption{\label{fig:plot103B}  Data Analysis, Snippet \texttt{lp3}, 5 years programming experience. $\hat{F}^{(1)}_j$ - dashed line; $\hat{F}^{(2)}_j$ - dotted line; $\hat{F}^{(3)}_j$ - continuous line. Red - incorrect answer; green - correct answer, blue - marginal probability.}
\end{figure}

\begin{figure}[H]
	\centering
	\makebox{\includegraphics[width=12cm,height=16cm]{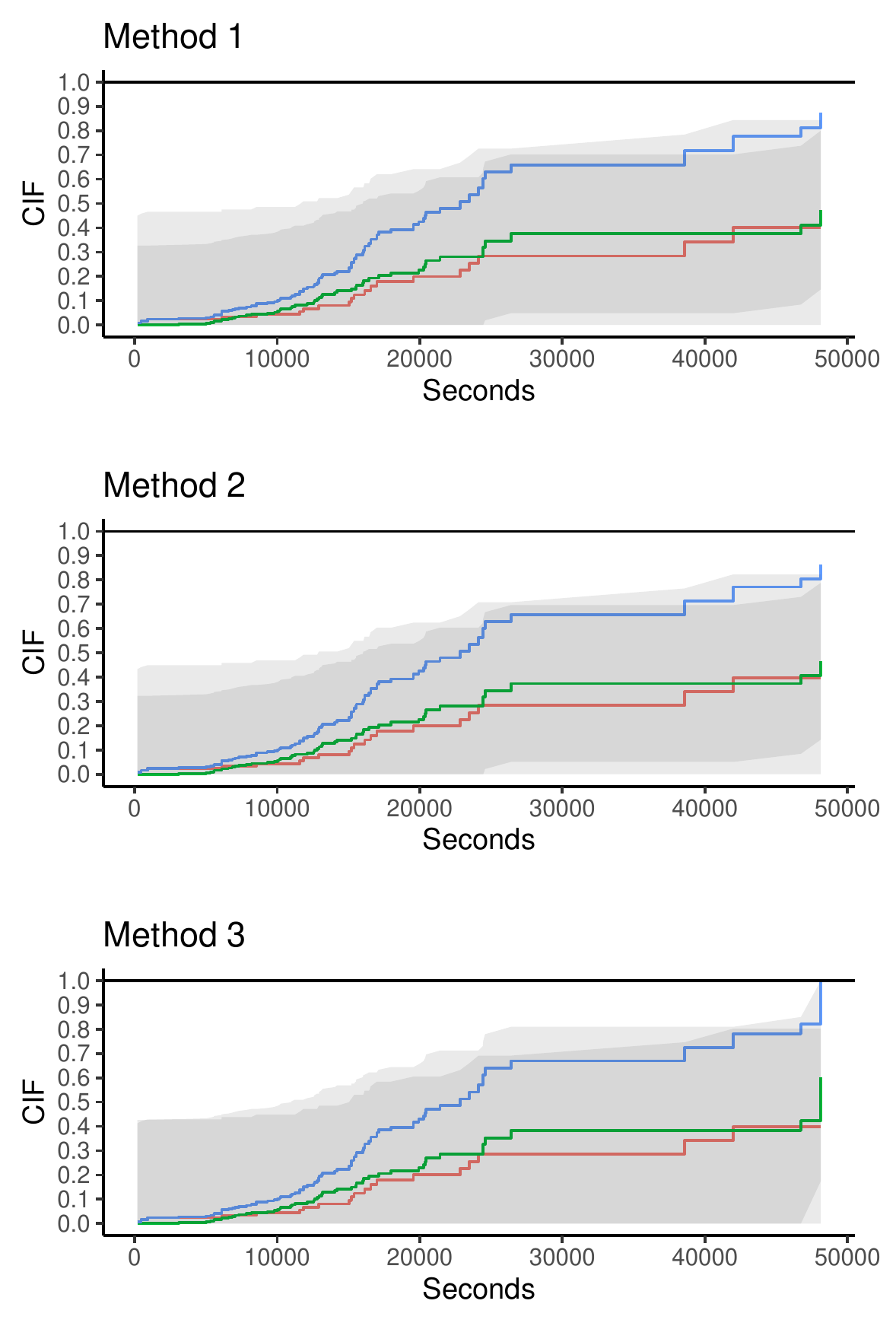}}
	\caption{\label{fig:plot103C}  Data Analysis, Snippet \texttt{lp3}, $\bz = (\mbox{snippet order}=1,\mbox{age}=35,\mbox{female},\mbox{YoE}=5)$.  Method 1 - $\hat{F}^{(1)}_j$; Method 2 - $\hat{F}^{(2)}_j$; Method 3 - $\hat{F}^{(3)}_j$. Red - incorrect answer; green - correct answer, blue - marginal probability.}
\end{figure}

\newpage

\vspace*{20mm}

\begin{center}
\Large
{\bf Web-based Supplementary Materials for \\
Revisiting the Cumulative Incidence Function With Competing Risks Data \\
by David M. Zucker and Malka Gorfine}
\end{center}

\newpage

\setlength{\oddsidemargin}{-0.8in}
\setlength{\evensidemargin}{0.0in}
\setlength{\topmargin}{-0.75in}
\setlength{\textheight}{9.25in}
\setlength{\textwidth}{8in}

\footnotesize
\setstretch{1}

\begin{center}
	Table S.1 \\
	Maximum Bias of CDF Estimates -- Uniformly Distributed Covariate \\[12pt]
	\begin{tabular}{|cccccc|ccc|ccc|}
		\hline
		& Hazard & Sample & Relative & Covariate &           & \multicolumn{3}{c|}{Max Bias Cause A CDF} & \multicolumn{3}{c|}{Max Bias Cause B CDF} \\
		\cline{7-12}         
		Scenario &  Shape &   Size &     Risk &     Value & Censoring & Method 1 & Method 2 & Method 3 &  Method 1 & Method 2 & Method 3 \\
		\hline
		1  & increasing  &    75  &    3  &   $-0.4$  &     0\%   &     0.0129   &       0.0052   &       0.0029   &       0.0114  &        0.0021   &       0.0032   \\
		2  & increasing  &   150  &    3  &   $-0.4$  &    50\%   &     0.0174   &       0.0039   &       0.0086   &       0.0025  &        0.0084   &       0.0050   \\
		3  & increasing  &    75  &    3  &  \ph0.0   &     0\%   &     0.0223   &       0.0031   &       0.0031   &       0.0110  &        0.0035   &       0.0033   \\
		4  & increasing  &   150  &    3  &  \ph0.0   &    50\%   &     0.0286   &       0.0046   &       0.0043   &       0.0092  &        0.0036   &       0.0045   \\
		5  & increasing  &    75  &    3  &  \ph0.4   &     0\%   &     0.0152   &       0.0049   &       0.0044   &       0.0143  &        0.0039   &       0.0036   \\
		6  & increasing  &   150  &    3  &  \ph0.4   &    50\%   &     0.0174   &       0.0055   &       0.0065   &       0.0166  &        0.0047   &       0.0038   \\
		7  & increasing  &    75  &    6  &   $-0.4$  &     0\%   &     0.0158   &       0.0050   &       0.0047   &       0.0125  &        0.0019   &       0.0041   \\
		8  & increasing  &   150  &    6  &   $-0.4$  &    50\%   &     0.0168   &       0.0051   &       0.0108   &       0.0024  &        0.0122   &       0.0080   \\
		9  & increasing  &    75  &    6  &  \ph0.0   &     0\%   &     0.0240   &       0.0035   &       0.0038   &       0.0124  &        0.0041   &       0.0036   \\
		10  & increasing  &   150  &    6  &  \ph0.0   &    50\%   &     0.0342   &       0.0052   &       0.0041   &       0.0126  &        0.0037   &       0.0046   \\
		11  & increasing  &    75  &    6  &  \ph0.4   &     0\%   &     0.0150   &       0.0047   &       0.0052   &       0.0149  &        0.0047   &       0.0041   \\
		12  & increasing  &   150  &    6  &  \ph0.4   &    50\%   &     0.0165   &       0.0054   &       0.0063   &       0.0162  &        0.0047   &       0.0041   \\
		\hline
		13  &  decreasing  &   75  &    3  &   $-0.4$  &     0\%   &     0.0105   &       0.0081   &       0.0052   &       0.0142  &        0.0072   &       0.0082   \\
		14  &  decreasing  &  150  &    3  &   $-0.4$  &    50\%   &     0.0092   &       0.0148   &       0.0048   &       0.0123  &        0.0151   &       0.0089   \\
		15  &  decreasing  &   75  &    3  &  \ph0.0   &     0\%   &     0.0225   &       0.0037   &       0.0034   &       0.0108  &        0.0020   &       0.0019   \\
		16  &  decreasing  &  150  &    3  &  \ph0.0   &    50\%   &     0.0147   &       0.0051   &       0.0047   &       0.0033  &        0.0091   &       0.0116   \\
		17  &  decreasing  &   75  &    3  &  \ph0.4   &     0\%   &     0.0187   &       0.0016   &       0.0027   &       0.0108  &        0.0009   &       0.0013   \\
		18  &  decreasing  &  150  &    3  &  \ph0.4   &    50\%   &     0.0159   &       0.0044   &       0.0097   &       0.0108  &        0.0066   &       0.0060   \\
		19  &  decreasing  &   75  &    6  &   $-0.4$  &     0\%   &     0.0109   &       0.0100   &       0.0055   &       0.0142  &        0.0055   &       0.0071   \\
		20  &  decreasing  &  150  &    6  &   $-0.4$  &    50\%   &     0.0209   &       0.0249   &       0.0052   &       0.0153  &        0.0175   &       0.0088   \\
		21  &  decreasing  &   75  &    6  &  \ph0.0   &     0\%   &     0.0238   &       0.0030   &       0.0024   &       0.0125  &        0.0017   &       0.0017   \\
		22  &  decreasing  &  150  &    6  &  \ph0.0   &    50\%   &     0.0149   &       0.0057   &       0.0104   &       0.0024  &        0.0088   &       0.0133   \\
		23  &  decreasing  &   75  &    6  &  \ph0.4   &     0\%   &     0.0185   &       0.0025   &       0.0038   &       0.0112  &        0.0024   &       0.0030   \\ 24  &  decreasing  &  150  &    6  &  \ph0.4   &    50\%   &     0.0200   &       0.0030   &       0.0077   &       0.0122  &        0.0049   &       0.0044   \\
		\hline
		25  & up \& down   &    75  &    3  &   $-0.4$  &     0\%  &     0.0107   &       0.0079   &       0.0049   &       0.0146  &        0.0079   &       0.0090   \\
		26  & up \& down   &   150  &    3  &   $-0.4$  &    50\%  &     0.0074   &       0.0122   &       0.0051   &       0.0018  &        0.0074   &       0.0029   \\
		27  & up \& down   &    75  &    3  &  \ph0.0   &     0\%  &     0.0223   &       0.0037   &       0.0034   &       0.0109  &        0.0021   &       0.0020   \\
		28  & up \& down   &   150  &    3  &  \ph0.0   &    50\%  &     0.0229   &       0.0054   &       0.0046   &       0.0077  &        0.0039   &       0.0054   \\
		29  & up \& down   &    75  &    3  &  \ph0.4   &     0\%  &     0.0187   &       0.0024   &       0.0023   &       0.0107  &        0.0012   &       0.0011   \\
		30  & up \& down   &   150  &    3  &  \ph0.4   &    50\%  &     0.0169   &       0.0042   &       0.0051   &       0.0153  &        0.0080   &       0.0074   \\
		31  & up \& down   &    75  &    6  &   $-0.4$  &     0\%  &     0.0132   &       0.0081   &       0.0048   &       0.0153  &        0.0070   &       0.0089   \\
		32  & up \& down   &   150  &    6  &   $-0.4$  &    50\%  &     0.0071   &       0.0176   &       0.0087   &       0.0053  &        0.0126   &       0.0063   \\
		33  & up \& down   &    75  &    6  &  \ph0.0   &     0\%  &     0.0239   &       0.0039   &       0.0033   &       0.0125  &        0.0021   &       0.0019   \\
		34  & up \& down   &   150  &    6  &  \ph0.0   &    50\%  &     0.0295   &       0.0067   &       0.0058   &       0.0092  &        0.0048   &       0.0070   \\
		35  & up \& down   &    75  &    6  &  \ph0.4   &     0\%  &     0.0186   &       0.0023   &       0.0029   &       0.0112  &        0.0012   &       0.0017   \\
		36  & up \& down   &   150  &     6  & \ph0.4   &    50\%  &     0.0175   &       0.0038   &       0.0047   &       0.0145  &        0.0075   &       0.0067   \\
		\hline
	\end{tabular}
\end{center}

\newpage

\begin{center}
	Table S.2 \\
	End-of-Study SD of CDF Estimates -- Uniformly Distributed Covariate \\[12pt]
	\begin{tabular}{|cccccc|ccc|ccc|}
		\hline
		& Hazard & Sample & Relative & Covariate &           & \multicolumn{3}{c|}{Final SD Cause A CDF} & \multicolumn{3}{c|}{Final SD Cause B CDF} \\
		\cline{7-12}         
		Scenario &  Shape &   Size &     Risk &     Value & Censoring & Method 1 & Method 2 & Method 3 &  Method 1 & Method 2 & Method 3 \\
		\hline
		1  & increasing  &    75  &    3  &   $-0.4$  &     0\%   &      0.0973   &     0.0937  &     0.0916   &     0.0967    &    0.0936   &     0.0917  \\
		2  & increasing  &   150  &    3  &   $-0.4$  &    50\%   &      0.1076   &     0.1024  &     0.1001   &     0.1055    &    0.1010   &     0.0989  \\
		3  & increasing  &    75  &    3  &  \ph0.0   &     0\%   &      0.0596   &     0.0577  &     0.0576   &     0.0595    &    0.0577   &     0.0576  \\
		4  & increasing  &   150  &    3  &  \ph0.0   &    50\%   &      0.0638   &     0.0596  &     0.0589   &     0.0621    &    0.0588   &     0.0584  \\
		5  & increasing  &    75  &    3  &  \ph0.4   &     0\%   &      0.1009   &     0.0981  &     0.0967   &     0.1014    &    0.0982   &     0.0967  \\
		6  & increasing  &   150  &    3  &  \ph0.4   &    50\%   &      0.0956   &     0.0920  &     0.0901   &     0.0956    &    0.0920   &     0.0901  \\
		7  & increasing  &    75  &    6  &   $-0.4$  &     0\%   &      0.0996   &     0.0948  &     0.0928   &     0.0988    &    0.0947   &     0.0931  \\
		8  & increasing  &   150  &    6  &   $-0.4$  &    50\%   &      0.1142   &     0.1071  &     0.1057   &     0.1113    &    0.1058   &     0.1042  \\
		9  & increasing  &    75  &    6  &  \ph0.0   &     0\%   &      0.0598   &     0.0581  &     0.0580   &     0.0598    &    0.0581   &     0.0580  \\
		10  & increasing  &   150  &    6  &  \ph0.0   &    50\%   &      0.0653   &     0.0607  &     0.0599   &     0.0640    &    0.0601   &     0.0596  \\
		11  & increasing  &    75  &    6  &  \ph0.4   &     0\%   &      0.1017   &     0.0988  &     0.0977   &     0.1020    &    0.0988   &     0.0977  \\
		12  & increasing  &   150  &    6  &  \ph0.4   &    50\%   &      0.0943   &     0.0908  &     0.0894   &     0.0943    &    0.0909   &     0.0894  \\
		\hline
		13  &  decreasing  &   75  &    3  &   $-0.4$  &     0\%   &      0.0956   &     0.0926  &     0.0905   &     0.0958    &    0.0927   &     0.0909  \\
		14  &  decreasing  &  150  &    3  &   $-0.4$  &    50\%   &      0.1333   &     0.1290  &     0.1386   &     0.1095    &    0.1068   &     0.1117  \\
		15  &  decreasing  &   75  &    3  &  \ph0.0   &     0\%   &      0.0552   &     0.0538  &     0.0536   &     0.0555    &    0.0539   &     0.0538  \\
		16  &  decreasing  &  150  &    3  &  \ph0.0   &    50\%   &      0.1037   &     0.0961  &     0.0864   &     0.0843    &    0.0788   &     0.0728  \\
		17  &  decreasing  &   75  &    3  &  \ph0.4   &     0\%   &      0.0971   &     0.0943  &     0.0928   &     0.0971    &    0.0943   &     0.0929  \\
		18  &  decreasing  &  150  &    3  &  \ph0.4   &    50\%   &      0.1067   &     0.0993  &     0.0908   &     0.0976    &    0.0920   &     0.0867  \\
		19  &  decreasing  &   75  &    6  &   $-0.4$  &     0\%   &      0.0976   &     0.0935  &     0.0917   &     0.0967    &    0.0930   &     0.0911  \\
		20  &  decreasing  &  150  &    6  &   $-0.4$  &    50\%   &      0.1515   &     0.1475  &     0.1634   &     0.1191    &    0.1163   &     0.1234  \\
		21  &  decreasing  &   75  &    6  &  \ph0.0   &     0\%   &      0.0564   &     0.0550  &     0.0548   &     0.0563    &    0.0550   &     0.0548  \\
		22  &  decreasing  &  150  &    6  &  \ph0.0   &    50\%   &      0.1225   &     0.1126  &     0.0901   &     0.0962    &    0.0890   &     0.0749  \\
		23  &  decreasing  &   75  &    6  &  \ph0.4   &     0\%   &      0.0978   &     0.0949  &     0.0938   &     0.0979    &    0.0949   &     0.0938  \\
		24  &  decreasing  &  150  &    6  &  \ph0.4   &    50\%   &      0.0973   &     0.0912  &     0.0845   &     0.0909    &    0.0865   &     0.0827  \\
		\hline 
		25  & up \& down   &    75  &    3  &   $-0.4$  &     0\%  &      0.0955   &     0.0925  &     0.0904   &     0.0955    &    0.0924   &     0.0905  \\
		26  & up \& down   &   150  &    3  &   $-0.4$  &    50\%  &      0.1095   &     0.1041  &     0.1033   &     0.1061    &    0.1017   &     0.1006  \\
		27  & up \& down   &    75  &    3  &  \ph0.0   &     0\%  &      0.0554   &     0.0539  &     0.0537   &     0.0553    &    0.0539   &     0.0537  \\
		28  & up \& down   &   150  &    3  &  \ph0.0   &    50\%  &      0.0660   &     0.0612  &     0.0602   &     0.0645    &    0.0605   &     0.0592  \\
		29  & up \& down   &    75  &    3  &  \ph0.4   &     0\%  &      0.0972   &     0.0943  &     0.0929   &     0.0972    &    0.0943   &     0.0929  \\
		30  & up \& down   &   150  &    3  &  \ph0.4   &    50\%  &      0.0961   &     0.0926  &     0.0908   &     0.0967    &    0.0929   &     0.0907  \\
		31  & up \& down   &    75  &    6  &   $-0.4$  &     0\%  &      0.0962   &     0.0921  &     0.0902   &     0.0969    &    0.0928   &     0.0908  \\
		32  & up \& down   &   150  &    6  &   $-0.4$  &    50\%  &      0.1220   &     0.1152  &     0.1150   &     0.1102    &    0.1053   &     0.1065  \\
		33  & up \& down   &    75  &    6  &  \ph0.0   &     0\%  &      0.0562   &     0.0549  &     0.0546   &     0.0563    &    0.0549   &     0.0547  \\
		34  & up \& down   &   150  &    6  &  \ph0.0   &    50\%  &      0.0683   &     0.0627  &     0.0610   &     0.0663    &    0.0617   &     0.0597  \\
		35  & up \& down   &    75  &    6  &  \ph0.4   &     0\%  &      0.0975   &     0.0946  &     0.0935   &     0.0976    &    0.0946   &     0.0935  \\
		36  & up \& down   &   150  &    6  &  \ph0.4   &    50\%  &      0.0942   &     0.0909  &     0.0896   &     0.0945    &    0.0910   &     0.0896  \\
		\hline
	\end{tabular}
\end{center}

\newpage

\begin{center}
	Table S.3 \\
	Empirical Coverage Rates of 95\% Confidence Bands -- Uniformly Distributed Covariate \\[12pt]
	\begin{tabular}{|cccccc|ccc|ccc|}
		\hline
		& Hazard & Sample & Relative & Covariate &           & \multicolumn{3}{c|}{Coverage Rate Cause A} & \multicolumn{3}{c|}{Coverage Rate Cause B CDF} \\
		\cline{7-12}
		Scenario &  Shape &   Size &     Risk &     Value & Censoring & Method 1 & Method 2 & Method 3 &  Method 1 & Method 2 & Method 3 \\
		\hline
		1  & increasing  &    75  &    3  &   $-0.4$  &     0\%   &    0.921  &   0.928 &    0.979  &   0.914  &   0.910  &   0.963   \\
		2  & increasing  &   150  &    3  &   $-0.4$  &    50\%   &    0.892  &   0.895 &    0.985  &   0.885  &   0.885  &   0.951   \\ 
		3  & increasing  &    75  &    3  &  \ph0.0   &     0\%   &    0.939  &   0.937 &    0.981  &   0.924  &   0.914  &   0.961   \\ 
		4  & increasing  &   150  &    3  &  \ph0.0   &    50\%   &    0.934  &   0.933 &    0.996  &   0.938  &   0.935  &   0.976   \\ 
		5  & increasing  &    75  &    3  &  \ph0.4   &     0\%   &    0.938  &   0.934 &    0.961  &   0.935  &   0.921  &   0.947   \\ 
		6  & increasing  &   150  &    3  &  \ph0.4   &    50\%   &    0.935  &   0.930 &    0.976  &   0.938  &   0.923  &   0.946   \\ 
		7  & increasing  &    75  &    6  &   $-0.4$  &     0\%   &    0.915  &   0.924 &    0.981  &   0.906  &   0.897  &   0.960   \\ 
		8  & increasing  &   150  &    6  &   $-0.4$  &    50\%   &    0.863  &   0.876 &    0.986  &   0.858  &   0.864  &   0.951   \\ 
		9  & increasing  &    75  &    6  &  \ph0.0   &     0\%   &    0.943  &   0.933 &    0.974  &   0.937  &   0.921  &   0.953   \\ 
		10  & increasing  &   150  &    6  &  \ph0.0   &    50\%   &    0.936  &   0.942 &    0.990  &   0.948  &   0.939  &   0.968   \\ 
		11  & increasing  &    75  &    6  &  \ph0.4   &     0\%   &    0.937  &   0.929 &    0.964  &   0.931  &   0.915  &   0.939   \\ 
		12  & increasing  &   150  &    6  &  \ph0.4   &    50\%   &    0.944  &   0.941 &    0.977  &   0.940  &   0.933  &   0.953   \\ 
		\hline                                                                                                                            
		13  &  decreasing  &   75  &    3  &   $-0.4$  &     0\%   &    0.917  &   0.918 &    0.990  &   0.920  &   0.916  &   0.962   \\ 
		14  &  decreasing  &  150  &    3  &   $-0.4$  &    50\%   &    0.816  &   0.826 &    0.963  &   0.851  &   0.855  &   0.916   \\ 
		15  &  decreasing  &   75  &    3  &  \ph0.0   &     0\%   &    0.949  &   0.937 &    0.992  &   0.955  &   0.944  &   0.981   \\ 
		16  &  decreasing  &  150  &    3  &  \ph0.0   &    50\%   &    0.843  &   0.859 &    0.992  &   0.889  &   0.890  &   0.941   \\ 
		17  &  decreasing  &   75  &    3  &  \ph0.4   &     0\%   &    0.944  &   0.937 &    0.966  &   0.934  &   0.925  &   0.950   \\ 
		18  &  decreasing  &  150  &    3  &  \ph0.4   &    50\%   &    0.936  &   0.939 &    0.977  &   0.942  &   0.933  &   0.957   \\ 
		19  &  decreasing  &   75  &    6  &   $-0.4$  &     0\%   &    0.913  &   0.918 &    0.993  &   0.908  &   0.910  &   0.968   \\ 
		20  &  decreasing  &  150  &    6  &   $-0.4$  &    50\%   &    0.778  &   0.785 &    0.946  &   0.820  &   0.821  &   0.914   \\ 
		21  &  decreasing  &   75  &    6  &  \ph0.0   &     0\%   &    0.950  &   0.945 &    0.985  &   0.947  &   0.941  &   0.971   \\ 
		22  &  decreasing  &  150  &    6  &  \ph0.0   &    50\%   &    0.837  &   0.860 &    0.985  &   0.881  &   0.888  &   0.944   \\ 
		23  &  decreasing  &   75  &    6  &  \ph0.4   &     0\%   &    0.949  &   0.942 &    0.972  &   0.934  &   0.920  &   0.951   \\ 
		24  &  decreasing  &  150  &    6  &  \ph0.4   &    50\%   &    0.937  &   0.938 &    0.975  &   0.951  &   0.934  &   0.953   \\ 
		\hline                                                                                                                         
		25  & up \& down   &    75  &    3  &   $-0.4$  &     0\%  &    0.924  &   0.923 &    0.993  &   0.920  &   0.922  &   0.965   \\ 
		26  & up \& down   &   150  &    3  &   $-0.4$  &    50\%  &    0.885  &   0.888 &    0.979  &   0.885  &   0.888  &   0.950   \\ 
		27  & up \& down   &    75  &    3  &  \ph0.0   &     0\%  &    0.953  &   0.940 &    0.991  &   0.959  &   0.949  &   0.981   \\ 
		28  & up \& down   &   150  &    3  &  \ph0.0   &    50\%  &    0.923  &   0.925 &    0.995  &   0.916  &   0.918  &   0.964   \\ 
		29  & up \& down   &    75  &    3  &  \ph0.4   &     0\%  &    0.940  &   0.937 &    0.965  &   0.934  &   0.925  &   0.949   \\ 
		30  & up \& down   &   150  &    3  &  \ph0.4   &    50\%  &    0.941  &   0.933 &    0.977  &   0.933  &   0.925  &   0.949   \\ 
		31  & up \& down   &    75  &    6  &   $-0.4$  &     0\%  &    0.915  &   0.924 &    0.993  &   0.911  &   0.913  &   0.967   \\ 
		32  & up \& down   &   150  &    6  &   $-0.4$  &    50\%  &    0.853  &   0.861 &    0.977  &   0.859  &   0.860  &   0.948   \\ 
		33  & up \& down   &    75  &    6  &  \ph0.0   &     0\%  &    0.955  &   0.954 &    0.987  &   0.952  &   0.946  &   0.976   \\ 
		34  & up \& down   &   150  &    6  &  \ph0.0   &    50\%  &    0.928  &   0.928 &    0.994  &   0.941  &   0.935  &   0.964   \\ 
		35  & up \& down   &    75  &    6  &  \ph0.4   &     0\%  &    0.949  &   0.944 &    0.972  &   0.936  &   0.916  &   0.946   \\ 
		36  & up \& down   &   150  &    6  &  \ph0.4   &    50\%  &    0.945  &   0.935 &    0.981  &   0.940  &   0.925  &   0.954   \\ 
		\hline
	\end{tabular}
\end{center}

\newpage

\begin{center}
	Table S.4 \\
	Half-Width of 95\% Confidence Bands -- Uniformly Distributed Covariate \\[12pt]
	\begin{tabular}{|cccccc|ccc|ccc|}
		\hline
		& Hazard & Sample & Relative & Covariate &           & \multicolumn{3}{c|}{Half-Width Cause A} & \multicolumn{3}{c|}{Half-Width Cause B} \\
		\cline{7-12}
		Scenario &  Shape &   Size &     Risk &     Value & Censoring & Method 1 & Method 2 & Method 3 &  Method 1 & Method 2 & Method 3 \\
		\hline
		1  & increasing  &    75  &    3  &   $-0.4$  &     0\%   &    0.2170 &  0.2117 &  0.2733  & 0.1949 &  0.1884 &  0.2391   \\
		2  & increasing  &   150  &    3  &   $-0.4$  &    50\%   &    0.2218 &  0.2162 &  0.3327  & 0.1953 &  0.1894 &  0.2636   \\ 
		3  & increasing  &    75  &    3  &  \ph0.0   &     0\%   &    0.1529 &  0.1473 &  0.1770  & 0.1316 &  0.1265 &  0.1414   \\ 
		4  & increasing  &   150  &    3  &  \ph0.0   &    50\%   &    0.1573 &  0.1492 &  0.2129  & 0.1357 &  0.1292 &  0.1589   \\ 
		5  & increasing  &    75  &    3  &  \ph0.4   &     0\%   &    0.2190 &  0.2113 &  0.2396  & 0.2015 &  0.1899 &  0.2065   \\ 
		6  & increasing  &   150  &    3  &  \ph0.4   &    50\%   &    0.2145 &  0.2052 &  0.2453  & 0.2007 &  0.1872 &  0.2082   \\ 
		7  & increasing  &    75  &    6  &   $-0.4$  &     0\%   &    0.2193 &  0.2140 &  0.2983  & 0.1943 &  0.1885 &  0.2544   \\ 
		8  & increasing  &   150  &    6  &   $-0.4$  &    50\%   &    0.2262 &  0.2206 &  0.3749  & 0.1946 &  0.1899 &  0.2867   \\ 
		9  & increasing  &    75  &    6  &  \ph0.0   &     0\%   &    0.1576 &  0.1512 &  0.1794  & 0.1359 &  0.1297 &  0.1426   \\ 
		10  & increasing  &   150  &    6  &  \ph0.0   &    50\%   &    0.1655 &  0.1549 &  0.2118  & 0.1438 &  0.1345 &  0.1590   \\ 
		11  & increasing  &    75  &    6  &  \ph0.4   &     0\%   &    0.2235 &  0.2151 &  0.2493  & 0.2034 &  0.1916 &  0.2104   \\ 
		12  & increasing  &   150  &    6  &  \ph0.4   &    50\%   &    0.2128 &  0.2034 &  0.2438  & 0.1980 &  0.1845 &  0.2054   \\ 
		\hline                                                                                                                        
		13  &  decreasing  &   75  &    3  &   $-0.4$  &     0\%   &    0.2179 &  0.2127 &  0.2764  & 0.1960 &  0.1898 &  0.2376   \\ 
		14  &  decreasing  &  150  &    3  &   $-0.4$  &    50\%   &    0.2124 &  0.2083 &  0.3410  & 0.1756 &  0.1725 &  0.2439   \\ 
		15  &  decreasing  &   75  &    3  &  \ph0.0   &     0\%   &    0.1525 &  0.1467 &  0.1767  & 0.1311 &  0.1263 &  0.1408   \\ 
		16  &  decreasing  &  150  &    3  &  \ph0.0   &    50\%   &    0.1695 &  0.1625 &  0.2464  & 0.1386 &  0.1339 &  0.1715   \\ 
		17  &  decreasing  &   75  &    3  &  \ph0.4   &     0\%   &    0.2175 &  0.2091 &  0.2379  & 0.1992 &  0.1880 &  0.2035   \\ 
		18  &  decreasing  &  150  &    3  &  \ph0.4   &    50\%   &    0.2156 &  0.2053 &  0.2390  & 0.1990 &  0.1866 &  0.1983   \\ 
		19  &  decreasing  &   75  &    6  &   $-0.4$  &     0\%   &    0.2214 &  0.2161 &  0.3020  & 0.1968 &  0.1909 &  0.2571   \\ 
		20  &  decreasing  &  150  &    6  &   $-0.4$  &    50\%   &    0.2057 &  0.2028 &  0.3658  & 0.1674 &  0.1654 &  0.2546   \\ 
		21  &  decreasing  &   75  &    6  &  \ph0.0   &     0\%   &    0.1572 &  0.1504 &  0.1787  & 0.1349 &  0.1291 &  0.1419   \\ 
		22  &  decreasing  &  150  &    6  &  \ph0.0   &    50\%   &    0.1843 &  0.1764 &  0.2470  & 0.1517 &  0.1461 &  0.1746   \\ 
		23  &  decreasing  &   75  &    6  &  \ph0.4   &     0\%   &    0.2208 &  0.2118 &  0.2467  & 0.2021 &  0.1902 &  0.2076   \\ 
		24  &  decreasing  &  150  &    6  &  \ph0.4   &    50\%   &    0.2101 &  0.1983 &  0.2271  & 0.1955 &  0.1810 &  0.1908   \\ 
		\hline                                                                                                                     
		25  & up \& down   &    75  &    3  &   $-0.4$  &     0\%  &    0.2167 &  0.2116 &  0.2758  & 0.1953 &  0.1891 &  0.2372   \\ 
		26  & up \& down   &   150  &    3  &   $-0.4$  &    50\%  &    0.2176 &  0.2122 &  0.3229  & 0.1914 &  0.1860 &  0.2632   \\ 
		27  & up \& down   &    75  &    3  &  \ph0.0   &     0\%  &    0.1517 &  0.1458 &  0.1757  & 0.1308 &  0.1259 &  0.1403   \\ 
		28  & up \& down   &   150  &    3  &  \ph0.0   &    50\%  &    0.1551 &  0.1471 &  0.2100  & 0.1345 &  0.1281 &  0.1592   \\ 
		29  & up \& down   &    75  &    3  &  \ph0.4   &     0\%  &    0.2170 &  0.2089 &  0.2375  & 0.1987 &  0.1876 &  0.2031   \\ 
		30  & up \& down   &   150  &    3  &  \ph0.4   &    50\%  &    0.2108 &  0.2016 &  0.2397  & 0.1977 &  0.1845 &  0.2040   \\ 
		31  & up \& down   &    75  &    6  &   $-0.4$  &     0\%  &    0.2199 &  0.2145 &  0.3000  & 0.1952 &  0.1894 &  0.2564   \\ 
		32  & up \& down   &   150  &    6  &   $-0.4$  &    50\%  &    0.2211 &  0.2161 &  0.3581  & 0.1888 &  0.1847 &  0.2799   \\ 
		33  & up \& down   &    75  &    6  &  \ph0.0   &     0\%  &    0.1567 &  0.1498 &  0.1778  & 0.1351 &  0.1291 &  0.1417   \\ 
		34  & up \& down   &   150  &    6  &  \ph0.0   &    50\%  &    0.1650 &  0.1544 &  0.2105  & 0.1434 &  0.1345 &  0.1579   \\ 
		35  & up \& down   &    75  &    6  &  \ph0.4   &     0\%  &    0.2211 &  0.2123 &  0.2474  & 0.2016 &  0.1898 &  0.2073   \\ 
		36  & up \& down   &   150  &    6  &  \ph0.4   &    50\%  &    0.2084 &  0.1989 &  0.2385  & 0.1942 &  0.1812 &  0.2006   \\ 
		\hline       
	\end{tabular}
\end{center}

\newpage

\begin{center}
	Table S.5 \\
	Maximum Bias of CDF Estimates -- Normally Distributed Covariate \\[12pt]
	\begin{tabular}{|cc|ccc|ccc|}
		\hline
		Relative & Covariate &  \multicolumn{3}{c|}{Max Bias Cause A CDF} & \multicolumn{3}{c|}{Max Bias Cause B CDF} \\
		\cline{3-8}
		Risk &     Value &  Method 1 & Method 2 & Method 3 &  Method 1 & Method 2 & Method 3 \\
		\hline
		3 &  $-1.68$      &         0.0797   &      0.0090   &      0.0130   &      0.0359   &      0.0035  &       0.0055   \\
		3 &  \ph0.00      &         0.0338   &      0.0052   &      0.0088   &      0.0171   &      0.0025  &       0.0039   \\
		3 &  \ph1.68      &         0.0334   &      0.0024   &      0.0068   &      0.0189   &      0.0057  &       0.0039   \\
		6 &  $-1.68$      &         0.1406   &      0.0344   &      0.0328   &      0.0760   &      0.0191  &       0.0185   \\
		6 &  \ph0.00      &         0.0425   &      0.0098   &      0.0159   &      0.0251   &      0.0053  &       0.0079   \\
		6 &  \ph1.68      &         0.0456   &      0.0056   &      0.0131   &      0.0308   &      0.0035  &       0.0058   \\
		\hline
	\end{tabular} 
\end{center}

\begin{center}
	Table S.6 \\
	End-of-Study SD of CDF Estimates -- Normally Distributed Covariate \\[12pt]
	\begin{tabular}{|cc|ccc|ccc|}
		\hline
		Relative & Covariate &  \multicolumn{3}{c|}{Max Bias Cause A CDF} & \multicolumn{3}{c|}{Max Bias Cause B CDF} \\
		\cline{3-8}
		Risk &     Value &  Method 1 & Method 2 & Method 3 &  Method 1 & Method 2 & Method 3 \\
		\hline      
		3 &  $-1.68$      &      0.1127   &     0.0959   &     0.0901    &    0.1028   &     0.0924  &      0.0901   \\
		3 &  \ph0.00      &      0.0754   &     0.0729   &     0.0721    &    0.0757   &     0.0729  &      0.0721   \\
		3 &  \ph1.68      &      0.0928   &     0.0890   &     0.0883    &    0.0933   &     0.0890  &      0.0883   \\
		6 &  $-1.68$      &      0.4492   &     0.2215   &     0.1142    &    0.1930   &     0.1428  &      0.1143   \\
		6 &  \ph0.00      &      0.0938   &     0.0886   &     0.0868    &    0.0940   &     0.0886  &      0.0868   \\
		6 &  \ph1.68      &      0.1074   &     0.1003   &     0.0979    &    0.1071   &     0.1003  &      0.0979   \\
		\hline  
	\end{tabular}
\end{center} 

\begin{center}
	Table S.7 \\
	Empirical Coverage Rates of 95\% Confidence Bands -- Normally Distributed Covariate \\[12pt]
	\begin{tabular}{|cc|ccc|ccc|}
		\hline
		Relative & Covariate &  \multicolumn{3}{c|}{Max Bias Cause A CDF} & \multicolumn{3}{c|}{Max Bias Cause B CDF} \\
		\cline{3-8}
		Risk &     Value &  Method 1 & Method 2 & Method 3 &  Method 1 & Method 2 & Method 3 \\
		\hline 
		3 &  $-1.68$      &     0.937  &   0.914 &    0.980  &   0.948  &   0.914 &    0.955  \\
		3 &  \ph0.00      &     0.944  &   0.934 &    0.981  &   0.950  &   0.930 &    0.961  \\
		3 &  \ph1.68      &     0.956  &   0.956 &    0.985  &   0.947  &   0.923 &    0.944  \\
		6 &  $-1.68$      &     0.921  &   0.921 &    0.984  &   0.931  &   0.919 &    0.958  \\
		6 &  \ph0.00      &     0.946  &   0.927 &    0.972  &   0.939  &   0.915 &    0.950  \\
		6 &  \ph1.68      &     0.962  &   0.949 &    0.982  &   0.958  &   0.931 &    0.947  \\
		\hline
	\end{tabular}
\end{center} 

\begin{center}
	Table S.8 \\
	Half-Width of 95\% Confidence Bands -- Normally Distributed Covariate \\[12pt]
	\begin{tabular}{|cc|ccc|ccc|}
		\hline
		Relative & Covariate &  \multicolumn{3}{c|}{Max Bias Cause A CDF} & \multicolumn{3}{c|}{Max Bias Cause B CDF} \\
		\cline{3-8}
		Risk &     Value &  Method 1 & Method 2 & Method 3 &  Method 1 & Method 2 & Method 3 \\
		\hline 
		3 &  $-1.68$      &    0.3095 &  0.2531 &  0.3071 &  0.2545 &  0.2114 &  0.2367  \\
		3 &  \ph0.00      &    0.2093 &  0.1957 &  0.2369 &  0.1797 &  0.1664 &  0.1831  \\
		3 &  \ph1.68      &    0.2402 &  0.2240 &  0.2782 &  0.2110 &  0.1931 &  0.2158  \\
		6 &  $-1.68$      &    0.5386 &  0.3757 &  0.4009 &  0.3898 &  0.3012 &  0.3086  \\
		6 &  \ph0.00      &    0.2491 &  0.2272 &  0.2782 &  0.2175 &  0.1951 &  0.2141  \\
		6 &  \ph1.68      &    0.2907 &  0.2631 &  0.3247 &  0.2605 &  0.2298 &  0.2502  \\
		\hline    
	\end{tabular}
\end{center} 

\newpage

\begin{center}
	Table S.9 \\
	Quantiles of Total CIF At Last Event Time -- Uniformly Distributed Covariate\\[12pt]
	\begin{tabular}{|ccccc|ccccc|ccccc|}
		\hline
		& Hazard & Sample & Relative & Covariate & \multicolumn{5}{c|}{Quantiles for Method 1} & \multicolumn{5}{c|}{Quantiles for Method 2} \\
		\cline{6-15}
		Scenario &  Shape &   Size &     Risk &     Value &
		0.01 &  0.10 &   0.50 &  0.90  & 0.99 &  0.01 &  0.10 &  0.50 &  0.90 &  0.99 \\
		\hline
		1  & increasing  &    75  &    3  &   $-0.4$  &     0.9891 &  1.0102 &  1.0265 &  1.0387 &  1.0544 &  0.9688 &  0.9849 &  0.9979 &  1.0036 &  1.0113  \\
		3  & increasing  &    75  &    3  &  \ph0.0   &     1.0223 &  1.0277 &  1.0345 &  1.0408 &  1.0490 &  0.9975 &  1.0000 &  1.0010 &  1.0038 &  1.0103  \\
		5  & increasing  &    75  &    3  &  \ph0.4   &     1.0264 &  1.0277 &  1.0291 &  1.0327 &  1.0387 &  0.9998 &  1.0000 &  1.0001 &  1.0010 &  1.0043  \\
		7  & increasing  &    75  &    6  &   $-0.4$  &     0.9763 &  1.0070 &  1.0328 &  1.0510 &  1.0722 &  0.9576 &  0.9818 &  0.9989 &  1.0078 &  1.0206  \\
		9  & increasing  &    75  &    6  &  \ph0.0   &     1.0275 &  1.0322 &  1.0368 &  1.0425 &  1.0499 &  0.9995 &  1.0000 &  1.0005 &  1.0025 &  1.0073  \\
		11  & increasing  &    75  &    6  &  \ph0.4   &     1.0275 &  1.0279 &  1.0292 &  1.0326 &  1.0378 &  1.0000 &  1.0000 &  1.0000 &  1.0001 &  1.0006  \\
		\hline
		13  &  decreasing  &   75  &    3  &   $-0.4$  &     0.9884 &  1.0099 &  1.0268 &  1.0382 &  1.0529 &  0.9678 &  0.9854 &  0.9982 &  1.0032 &  1.0101  \\
		15  &  decreasing  &   75  &    3  &  \ph0.0   &     1.0228 &  1.0274 &  1.0342 &  1.0407 &  1.0501 &  0.9974 &  1.0000 &  1.0010 &  1.0037 &  1.0087  \\
		17  &  decreasing  &   75  &    3  &  \ph0.4   &     1.0262 &  1.0277 &  1.0291 &  1.0324 &  1.0390 &  0.9995 &  1.0000 &  1.0001 &  1.0011 &  1.0041  \\
		19  &  decreasing  &   75  &    6  &   $-0.4$  &     0.9694 &  1.0002 &  1.0291 &  1.0483 &  1.0654 &  0.9513 &  0.9765 &  0.9971 &  1.0065 &  1.0178  \\
		21  &  decreasing  &   75  &    6  &  \ph0.0   &     1.0272 &  1.0319 &  1.0367 &  1.0419 &  1.0511 &  0.9995 &  1.0001 &  1.0005 &  1.0029 &  1.0074  \\
		23  &  decreasing  &   75  &    6  &  \ph0.4   &     1.0274 &  1.0279 &  1.0292 &  1.0323 &  1.0369 &  1.0000 &  1.0000 &  1.0000 &  1.0002 &  1.0012  \\
		\hline
		25  & up \& down   &    75  &   3  &   $-0.4$  &    0.9893  & 1.0099  & 1.0271  & 1.0385  & 1.0528  & 0.9681  & 0.9853  & 0.9982  & 1.0032  & 1.0098   \\
		27  & up \& down   &    75  &   3  &  \ph0.0   &    1.0227  & 1.0277  & 1.0344  & 1.0407  & 1.0495  & 0.9973  & 1.0000  & 1.0011  & 1.0037  & 1.0087   \\
		29  & up \& down   &    75  &   3  &  \ph0.4   &    1.0263  & 1.0278  & 1.0291  & 1.0324  & 1.0392  & 0.9996  & 1.0000  & 1.0001  & 1.0011  & 1.0040   \\
		31  & up \& down   &    75  &   6  &   $-0.4$  &    0.9767  & 1.0053  & 1.0318  & 1.0510  & 1.0675  & 0.9546  & 0.9792  & 0.9984  & 1.0072  & 1.0165   \\
		33  & up \& down   &    75  &   6  &  \ph0.0   &    1.0277  & 1.0324  & 1.0370  & 1.0420  & 1.0505  & 0.9997  & 1.0000  & 1.0005  & 1.0026  & 1.0066   \\
		35  & up \& down   &    75  &   6  &  \ph0.4   &    1.0275  & 1.0279  & 1.0293  & 1.0325  & 1.0375  & 1.0000  & 1.0000  & 1.0000  & 1.0002  & 1.0012   \\
		\hline
	\end{tabular}
\end{center}

\begin{center}
	Table S.10 \\
	Quantiles of Total CIF At Last Event Time -- Normally Distributed Covariate\\[12pt]
	\begin{tabular}{|cc|ccccc|ccccc|}
		\hline
		Relative & Covariate & \multicolumn{5}{c|}{Quantiles for Method 1} & \multicolumn{5}{c|}{Quantiles for Method 2} \\
		\cline{3-12}
		Risk &     Value &
		0.01 &  0.10 &   0.50 &  0.90  & 0.99 &  0.01 &  0.10 &  0.50 &  0.90 &  0.99 \\
		\hline
		3 &  $-1.68$      &      1.0561 &  1.0762  & 1.1023  & 1.1638 &  1.3994 &  0.9981 &  1.0000 &  1.0014 &  1.0170 &  1.0846  \\
		3 &  \ph0.00      &      1.0384 &  1.0429  & 1.0497  & 1.0598 &  1.0743 &  1.0000 &  1.0000 &  1.0000 &  1.0002 &  1.0017  \\
		3 &  \ph1.68      &      1.0388 &  1.0430  & 1.0510  & 1.0634 &  1.0769 &  1.0000 &  1.0000 &  1.0000 &  1.0000 &  1.0004  \\
		6 &  $-1.68$      &      0.9935 &  1.0893  & 1.1553  & 1.3269 &  1.9050 &  0.9503 &  1.0001 &  1.0063 &  1.0690 &  1.3984  \\
		6 &  \ph0.00      &      1.0468 &  1.0527  & 1.0651  & 1.0859 &  1.1114 &  1.0000 &  1.0000 &  1.0000 &  1.0004 &  1.0049  \\
		6 &  \ph1.68      &      1.0528 &  1.0595  & 1.0736  & 1.0951 &  1.1382 &  1.0000 &  1.0000 &  1.0000 &  1.0002 &  1.0045  \\
		\hline
	\end{tabular}
\end{center}

\end{document}